\documentclass[acmsmall,screen,9pt]{acmart}
\usepackage{soul}
\soulregister\cite7
\soulregister\ref7

\usepackage{array}
\usepackage{amssymb}
\usepackage{caption}
\usepackage{textcomp}
\usepackage{utfsym}
\usepackage{tabularx}
\usepackage{amsmath}
\usepackage{stfloats}
\usepackage{url}
\usepackage{verbatim}
\usepackage{graphicx}
\usepackage{float}
\usepackage{multirow}
\usepackage{booktabs}
\usepackage{makecell}
\usepackage{wrapfig}

\AtBeginDocument{%
  }

\setcopyright{acmcopyright}

\acmJournal{CSUR}

\begin{document}

\title{Recent Advances in 3D Object and Scene Generation: A Survey}


\author{Xiang Tang}
\affiliation{%
  \institution{Harbin Institute of Technology, Shenzhen}
  \city{Shenzhen}
  \country{P.R.China}}
\email{24B951063@stu.hit.edu.cn}

\author{Ruotong Li}
\authornote{Corresponding author}
\affiliation{%
  \institution{Pengcheng Laboratory}
  \city{Shenzhen}
  \country{P.R.China}}
\email{lirt@pcl.ac.cn}

\author{Xiaopeng Fan}
\authornote{Corresponding author}
\affiliation{%
  \institution{Harbin Institute of Technology}
  \city{Harbin}
  \country{P.R.China}}
\email{fxp@hit.edu.cn}

\renewcommand{\shortauthors}{Tang et al.}




\begin{abstract}

In recent years, the demand for 3D content has grown exponentially with the intelligent upgrade of interactive media, extended reality (XR), and Metaverse industries. In order to overcome the limitations of traditional manual modeling approaches, such as labor-intensive workflows and prolonged production cycles, revolutionary advances have been achieved through the convergence of novel 3D representation paradigms and artificial intelligence generative technologies. In this survey, we conduct a systematic review of the cutting-edge achievements in static 3D object and scene generation, as well as establish a comprehensive technical framework through systematic categorization. We start our analysis with mainstream 3D object representations. Subsequently, we delve into the technical pathways of 3D object generation based on four mainstream deep generative models: Variational Autoencoders, Generative Adversarial Networks, Autoregressive Models, and Diffusion Models. Regarding scene generation, we focus on three dominant paradigms: layout-guided generation, lifting based on 2D priors, and rule-driven modeling. Finally, we critically examine persistent challenges in 3D generation and propose potential research directions for future investigation. This survey aims to provide readers with a structured understanding of state-of-the-art 3D generation technologies while inspiring researchers to undertake more exploration in this domain. Project page: \href{https://github.com/xdlbw/Awesome-3D-Object-and-Scene-Generation}{Awesome-3D-Object-and-Scene-Generation}.

\end{abstract}

\maketitle

\section{Introduction}

Over the decades, automated content generation has evolved significantly. In the early years, rule-based modeling such as L-system \cite{lindenmayer1968mathematical} and procedural shape grammar \cite{parish2001procedural, muller2006procedural} showed their efficiency in creating objects and scenes with regular and repetitive structures. Although it could generate 3D content with complex geometry and texture detail rapidly, the rules and grammars were manually designed and difficult to generalize until the neural networks and deep learning methods revolutionized computer vision in the 2010s. Guo et al. \cite{guo2020inverse} employed deep learning to discover atomic structures, extracting rules from input images and converting them into L-systems to achieve inverse procedural modeling. CropCraft \cite{zhai2024cropcraft} optimized plant morphological parameters through inverse procedural modeling to generate mesh representations of crops from images. With the breakthroughs in deep learning technologies, generative artificial intelligence (AI) has made revolutionary progress in 2D content generation: text parsing and generation models represented by DeepSeek \cite{guo2025deepseek}, text-to-image technologies led by Imagen \cite{saharia2022photorealistic} and GPT-4o \cite{hurst2024gpt}, all demonstrate outstanding performance. Against the backdrop of rapid advancements in metadata, 3D content generation has gained widespread attention as a natural extension of 2D technology, yet its development faces multiple challenges. The increase in dimensionality complicates the effective integration of explicit 3D representations into neural network architectures. Simultaneously, novel rendering techniques based on implicit Neural Radiance Fields \cite{mildenhall2021nerf} struggle to directly adapt their generated content to traditional rasterized graphics pipelines. Furthermore, the scarcity of high-quality 3D asset datasets significantly hinders model training.

\begin{figure}
    \centering
    \vspace{-0.3cm}
    \includegraphics[width=\textwidth]{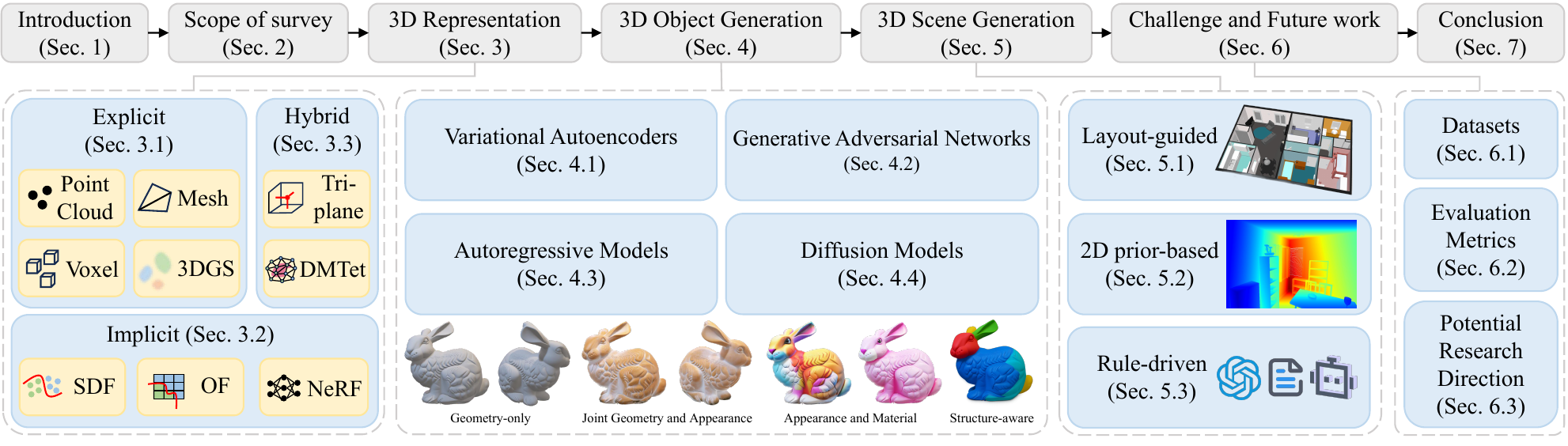}
    \vspace{-0.7cm}
    \caption{Structure of this survey.}
    \vspace{-0.6cm}
    \label{structure_of_this_survey}
\end{figure}

Despite these limitations, significant progress has been achieved in 3D generation, yielding several groundbreaking research outcomes. Among them, Point-E \cite{nichol2022point} constructed a million-scale 3D model-text paired dataset for training point cloud diffusion models, enabling text-to-3D point cloud generation. Several methods \cite{park2019deepsdf, luo2021surfgen, gao2022get3d, chen2023fantasia3d, lin2023magic3d, qian2023magic123, sun2023dreamcraft3d, liu2024sherpa3d} incorporated 3D implicit representations into deep generative models and subsequently extracted explicit meshes using algorithms \cite{lorensen1998marching, doi1991efficient}. Meanwhile, the proposal of novel representations such as 3D Gaussian Splatting \cite{kerbl20233d}, 3DShape2VecSet \cite{zhang20233dshape2vecset}, and PolyhedronNet \cite{yu2025polyhedronnet} has achieved key breakthroughs in real-time rendering efficiency, topological learning of explicit polyhedra, and vectorized compression for implicit generation, respectively. These representations have constructed geometric carriers that are more suitable for deep network learning and generation. DreamFusion \cite{poole2022dreamfusion} introduced an innovative paradigm leveraging 2D generative priors to supervise 3D representation optimization, establishing new directions for subsequent research. Concurrently, approaches such as \cite{ocal2024sceneteller, zhang2024towards, zhou2024gala3d, li2024dreamscene, gao2024graphdreamer, sun20233d, feng2025text} fully leveraged the powerful text parsing capabilities of large language models (LLMs) to extract scene features from natural language descriptions for layout construction. RGBD2 \cite{lei2023rgbd2} derived mesh representations of scenes from 2D image priors. The parametric generation framework proposed by Raistrick et al. \cite{raistrick2023infinite} successfully achieved infinite combinatorial generation of natural resources based on mathematical rules. Therefore, this study is dedicated to systematically analyzing the related research in the field of 3D content generation, summarizing its technical paradigms and categorizing them accordingly. As shown in Table \ref{summary_3d_generation_method}, we provide a structured overview of recent works, focusing on 3D representations, 3D object and scene generation methods.

\begin{table}
    \centering
    \vspace{-0.3cm}
    \caption{A summary of representative works in 3D generation, classified by "Target" and their generated "Content", where "Geo.", "App." and "Mat." represent Geometry, Appearance, and Material, respectively. The "GM" column indicates Generative Model, where "AR" denotes Autoregressive Models and "DM" represents Diffusion Models. The "3D Rep." column specifies the 3D Representation format of generated contents. The "Opt." column describes the Optimization domain or guidance strategy employed. Additionally, we present each method's required Input and whether it supports Editability.}
    \vspace{-0.3cm}
    \resizebox{\textwidth}{!}{
    \scriptsize
    \begin{tabular}{@{}lccccccc@{}}
    \toprule
    \textbf{Methods} & \textbf{Target} & \textbf{Content} & \textbf{GM} & \textbf{3D Rep.} & \textbf{Opt.} & \textbf{Input} & \textbf{Editability} \\
    \midrule
    DeepSDF \cite{park2019deepsdf} & Object & Geo. & VAE & SDF & 3D & Uncon. & \usym{2717} \\
    SetVAE \cite{kim2021setvae} & Object & Geo. & VAE & Point Cloud & 3D & Uncon. & \usym{2717} \\
    SurfGen \cite{luo2021surfgen} & Object & Geo. & GAN & SDF & 3D & Uncon. & \usym{2717} \\
    AutoSDF \cite{mittal2022autosdf} & Object & Geo. & AR  & SDF & 3D & Uncon./Text & \usym{2717} \\
    MeshXL \cite{chen2024meshxl} & Object & Geo. & AR  & Mesh & 3D & Uncon./Text/Image & \usym{2717} \\
    Hi3DGen \cite{ye2025hi3dgen} & Object & Geo. & DM  & Sparse Voxel & 3D & Image & \usym{2717} \\
    Direct3D-S2 \cite{wu2025direct3d} & Object & Geo. & DM  & Sparse Voxel & 3D & Image & \usym{2717} \\
    \midrule
    GET3D \cite{gao2022get3d} & Object & Joint Geo. \& App. & GAN & Tri-plane/DMTet & 3D & Uncon./Text & \usym{2714} \\
    Barthel et al. \cite{barthel2024gaussian} & Object & Joint Geo. \& App. & GAN & 3DGS & 3D & Uncon. & \usym{2714} \\
    SAR3D \cite{chen2025sar3d} & Object & Joint Geo. \& App. & AR & Tri-plane & 3D & Text/Image & \usym{2717} \\
    DreamFusion \cite{poole2022dreamfusion} & Object & Joint Geo. \& App. & DM  & NeRF & SDS & Text & \usym{2717} \\
    Yu et al. \cite{yu2023text} & Object & Joint Geo. \& App. & DM & NeRF/DMTet & CSD & Text & \usym{2714} \\
    LucidDreamer \cite{chung2023luciddreamer} & Object & Joint Geo. \& App. & DM & 3DGS & ISM & Text & \usym{2714} \\
    ProlificDreamer \cite{wang2023prolificdreamer} & Object & Joint Geo. \& App. & DM & NeRF/DMTet & VSD & Text & \usym{2717} \\
    Sculpt3D \cite{chen2024sculpt3d} & Object & Joint Geo. \& App. & DM & NeRF & VSD & Text & \usym{2717} \\
    ScaleDreamer \cite{ma2024scaledreamer} & Object & Joint Geo. \& App. & DM & NeRF & ASD & Text & \usym{2717} \\
    DreamControl \cite{huang2024dreamcontrol} & Object & Joint Geo. \& App. & DM & NeRF & SDS & Text & \usym{2717} \\
    Shap-E \cite{jun2023shap} & Object & Joint Geo. \& App. & DM & NeRF & 3D & Text/Image & \usym{2717} \\
    \midrule
    MaPa \cite{zhang2024mapa} & Object & App. \& Mat. & DM & - & 2D & Text/Image + Mesh & \usym{2714} \\
    TexGaussian \cite{xiong2024texgaussian} & Object & App. \& Mat. & DM & 3DGS & 3D & Uncon./Text & \usym{2717} \\
    MaterialMVP \cite{he2025materialmvp} & Object & App. \& Mat. & DM & - & 2D & Image+Mesh & \usym{2717} \\
    \midrule
    PQ-NET \cite{wu2020pq} & Object & Structure-aware & AR & SDF & 3D & Uncon./Image & \usym{2714} \\
    Part123 \cite{liu2024part123} & Object & Structure-aware & DM & SDF & 2D & Image & \usym{2714} \\
    PartCrafter \cite{lin2025partcrafter} & Object & Structure-aware & DM & Mesh & 3D & Image & \usym{2714} \\
    \midrule
    GRAINS \cite{li2019grains} & Scene & Indoor & VAE & Mesh & 3D & Uncon. & \usym{2714} \\
    PERF \cite{wang2024perf} & Scene & Indoor & DM & NeRF & 2D & Image & \usym{2714} \\
    InstructScene \cite{lin2024instructscene} & Scene & Indoor & DM & Mesh & 3D & Text & \usym{2717} \\
    MMGDreamer \cite{yang2025mmgdreamer} & Scene & Indoor & DM & SDF & 3D & Text/Image & \usym{2717} \\
    SceneCraft \cite{yang2024scenecraft} & Scene & Indoor & DM & NeRF & 2D & Text & \usym{2717} \\
    GenRC \cite{li2024genrc} & Scene & Indoor & DM & Mesh & 2D & Text/Image & \usym{2717} \\
    Dahnert et al. \cite{dahnert2024coherent} & Scene & Indoor & DM & 3DGS & 3D & Image & \usym{2717} \\
    \midrule
    DreamScene \cite{li2024dreamscene} & Scene & Cross & DM & 3DGS & CSD & Text & \usym{2714} \\
    WonderWorld \cite{yu2024wonderworld} & Scene & Cross & DM & 3DGS & 2D & Text & \usym{2717} \\
    CAST \cite{yao2025cast} & Scene & Cross & DM & Mesh & 2D \& 3D & Image & \usym{2717} \\
    DreamCube \cite{huang2025dreamcube} & Scene & Cross & DM & Mesh/3DGS & 2D & Text/Image & \usym{2717} \\
    Imaginarium \cite{zhu2025imaginarium} & Scene & Cross & DM & Mesh & 2D \& 3D & Text & \usym{2714} \\
    Gumin et al. \cite{gumin2025procedural} & Scene & Cross & LLM & Mesh & - & Text & \usym{2714} \\
    Scenethesis \cite{ling2025scenethesis} & Scene & Cross & VLM & Mesh & 3D & Text & \usym{2717} \\
    \midrule
    CityDreamer \cite{xie2024citydreamer} & Scene & Outdoor & GAN & NeRF & 2D & BEV Map & \usym{2714} \\
    Liu et al. \cite{liu2025controllable} & Scene & Outdoor & DM & Voxel & 2D \& 3D & BEV Map & \usym{2714} \\
    Sat2City \cite{hua2025sat2city} & Scene & Outdoor & DM & Voxel & 3D & Image & \usym{2717} \\
    BuildingBlick \cite{huang2025buildingblock} & Scene & Outdoor & DM & Mesh & 3D & Text & \usym{2714} \\
    Yo'City \cite{lu2025yo} & Scene & Outdoor & DM & Mesh & - & Text & \usym{2714} \\
    3D-GPT \cite{sun20233d} & Scene & Outdoor & LLM & Mesh & - & Text & \usym{2714} \\
    Proc-GS \cite{li2024proc} & Scene & Outdoor & LLM & 3DGS & - & Text/Image & \usym{2714} \\
    \bottomrule
    \end{tabular}}
    \vspace{-0.6cm}
    \label{summary_3d_generation_method}
\end{table}

As illustrated in Fig. \ref{structure_of_this_survey}, this survey first delineates its scope and related works in Section \ref{section2}. Section \ref{section3} introduces fundamental theories of 3D representations, analyzing their advantages, limitations, and integration with generative frameworks. Subsequently, we bifurcate 3D content generation into object-level and scene-level tasks. Section \ref{section4} focuses on four deep generative models, covering a complete technical pipeline from object shape exploration and texture synthesis to the understanding of functional structures. Expanding to scene generation in Section \ref{section5}, we classify methodologies into three categories based on their theory: scene synthesis guided by layouts or scene graphs, generation methods that directly extract scene representations from spatial information provided by 2D images, and rule-driven modeling with controllable details. Lastly, Section \ref{section6} identifies remaining challenges in the field and outlines future research directions. We aim to provide technical references for researchers and inspire subsequent works through this review.

Our principal contributions can be summarized as follows:

\begin{itemize}
\item[$\bullet$] We propose a novel taxonomy by decomposing 3D content generation into object and scene generation, systematically summarizing and categorizing their technical routes respectively.
\item[$\bullet$] This review comprehensively covers extensive literature spanning the past five years, emphasizing integration of the latest breakthroughs to comprehensively present technological development trajectories and cutting-edge dynamics.
\end{itemize}
\section{Scope of this survey}\label{section2}

This survey highlights recent advancements in 3D generation techniques for static object and scene generation tasks. Our focus lies on systematically categorizing and summarizing 3D generation paradigms and their application scenarios. The primary scope covers seminal papers published in top-tier computer vision and computer graphics conferences/journals from 2020 to 2025, supplemented by pre-prints from arXiv recently. Instead of exhaustive technical analysis, we curate representative studies to distill their core methodological frameworks, enabling readers to efficiently grasp technical trajectories and construct structured knowledge graphs. It should be noted that this survey explicitly excludes dynamic 3D content and human-related generation methods (e.g., avatar modeling, full-body reconstruction, and motion synthesis). 

Table \ref{survey_comparison} outlines the distinctions between our work and existing representative surveys in terms of scope and taxonomic classification. Diverging from early reviews that prioritized structured and procedural modeling \cite{chaudhuri2020learning}, or those primarily analyzing the compatibility between 3D representations and generative models \cite{shi2022deep}, our study focuses on the intrinsic evolution of deep generative techniques. In contrast to surveys limited to either scene understanding or generation \cite{patil2024advances, wen20253d}, we establish a unified generative perspective that encompasses both 3D objects and scenes. Regarding input modalities, while previous works often confined their scope to text-to-3D generation \cite{li2023generative, jiang2024survey}, this paper incorporates both image-driven and unconditional generation. For a broader discussion on audio-video cross-modality generation, readers may refer to \cite{foo2025ai}. Notably, serving as a significant complement to existing comprehensive surveys \cite{li2024advances, liu2024comprehensive}, we propose a taxonomy oriented towards the generation targets. This framework holistically covers complete technical pathways for 3D object and scene generation, while proposing novel insights into methodological evolution and application expansion.

\begin{table}
    \centering
    \vspace{-0.3cm}
    \caption{Comparison of the scope across different surveys. "Temporal Distribution" indicates the proportion of cited references published in 2023 or earlier, 2024, and 2025. The symbol \protect\usym{2757} denotes a brief mention.}
    \vspace{-0.3cm}
    \resizebox{\textwidth}{!}{
    \begin{tabular}{@{}cccccccccclccc@{}}
    \toprule

    \multirow{2}{*}{\centering \makecell{\textbf{Survey}}} &  
    \multicolumn{5}{c}{\textbf{Generative Target}} & 
    \multicolumn{3}{c}{\textbf{Generative Modality}} & 
    \multirow{2}{*}{\centering \makecell{\textbf{Application}}} &
    \multirow{2}{*}{\centering \makecell{\textbf{Taxonomy}}} &
    \multicolumn{3}{c}{\centering \makecell{\textbf{Temporal Distribution}}} \\
    \cmidrule(lr){2-6} \cmidrule(lr){7-9} \cmidrule(lr){12-14}
     & \textbf{Object} & \textbf{Scene} & \textbf{Texture} & \textbf{Human} & \textbf{Dynamic} & \textbf{Text} & \textbf{Image} & \textbf{Procedural} & & & \textbf{$\le$2023} & \textbf{2024} & \textbf{2025}\\

    \midrule
    \cite{chaudhuri2020learning} & 
    \usym{2714} & \usym{2714} & \usym{2717} & \usym{2717} & \usym{2717} & \usym{2717} & \usym{2717} & \usym{2714} & \usym{2714} & 
    Generative Models & 100\% & 0 & 0 \\
    \midrule
    \cite{shi2022deep} & 
    \usym{2714} & \usym{2714} & \usym{2717} & \usym{2717} & \usym{2717} & \usym{2717} & \usym{2714} & \usym{2717} & \usym{2714} & 
    Representation-Model Coupling & 100\% & 0 & 0 \\
    \midrule
    \cite{patil2024advances} & 
    \usym{2717} & \usym{2714} & \usym{2717} & \usym{2717} & \usym{2717} & \usym{2717} & \usym{2714} & \usym{2714} & \usym{2714} & 
    Scene Tasks & 100\% & 0 & 0 \\
    \midrule
    \cite{li2023generative} & 
    \usym{2714} & \usym{2714} & \usym{2714} & \usym{2714} & \usym{2717} & \usym{2714} & \usym{2717} & \usym{2717} & \usym{2714} & 
    Performance Optimization & 79\% & 21\% & 0 \\
    \midrule
    \cite{foo2025ai} & 
    \usym{2714} & \usym{2714} & \usym{2717} & \usym{2714} & \usym{2714} & \usym{2714} & \usym{2714} & \usym{2717} & \usym{2714} & 
    Multimodal Generation & 98.7\% & 1.3\% & 0 \\
    \midrule
    \cite{li2024advances} & 
    \usym{2714} & \usym{2714} & \usym{2757} & \usym{2714} & \usym{2717} & \usym{2714} & \usym{2714} & \usym{2714} & \usym{2714} & 
    Generation Methods & 99.6\% & 0.4\% & 0 \\
    \midrule
    \cite{liu2024comprehensive} & 
    \usym{2714} & \usym{2714} & \usym{2757} & \usym{2714} & \usym{2757} & \usym{2714} & \usym{2714} & \usym{2717} & \usym{2757} & 
    Generation Methods & 89.9\% & 10.1\% & 0 \\
    \midrule
    \cite{jiang2024survey} & 
    \usym{2714} & \usym{2717} & \usym{2717} & \usym{2717} & \usym{2717} & \usym{2714} & \usym{2717} & \usym{2717} & \usym{2757} & 
    Generation Methods & 80.2\% & 19.8\% & 0 \\
    \midrule
    \cite{wen20253d} & 
    \usym{2717} & \usym{2714} & \usym{2717} & \usym{2757} & \usym{2714} & \usym{2714} & \usym{2714} & \usym{2714} & \usym{2714} & 
    Generation Methods & 51.3\% & 31.5\% & 17.2\% \\
    \midrule
    Ours & 
    \usym{2714} & \usym{2714} & \usym{2714} & \usym{2717} & \usym{2717} & \usym{2714} & \usym{2714} & \usym{2714} & \usym{2714} & 
    Generation Methods and Targets & 41.6\% & 24.7\% & 33.7\% \\
    \bottomrule
    \end{tabular}}
    \vspace{-0.6cm}
    \label{survey_comparison}
\end{table}
\section{3D Representations}\label{section3}

In the field of 3D content generation, the choice of 3D model representation not only influences the generative method chosen but also affects the quality and efficiency of the generation algorithms. In this section, we go through 3D representations and categorize them based on mathematical expressions and geometric description mechanisms: explicit representation, implicit representation, and hybrid representation.

\subsection{Explicit Representations}

Typical explicit representations include meshes, point clouds, voxels, and the emerging 3D Gaussian Splatting. Such representations directly describe the spatial structure of 3D objects through discrete geometric elements or primitives. They support geometric editing and precise control of geometry details and are widely used in traditional computer graphics tasks such as modeling, rendering, and simulation.

\subsubsection{Mesh}

Polygonal mesh representation, as a fundamental paradigm for 3D geometric modeling, defines geometric shapes by explicitly encoding the 3D coordinates of vertices and their topological adjacency relationships, enabling compact and precise geometric descriptions. Thanks to its explicit parameterization, mesh models support efficient rendering operations and affine transformations (including translation, rotation, scaling, etc.). As an early attempt to generate object meshes, Pixel2Mesh \cite{wang2018pixel2mesh} employed graph convolutional networks (GCN) to process meshes by defining model vertices and edges as nodes and connections in the graph, directly generating 3D meshes from single images. Gkioxari et al. \cite{gkioxari2019mesh} leveraged GCN to iteratively optimize mesh vertex positions initialized from voxels. By fusing image features through vertex alignment, they enhanced detail accuracy and constructed a system to achieve joint 2D detection and 3D mesh extraction from real-world images.

\subsubsection{Point clouds}

Point clouds consist of unordered discrete sampling points distributed in three-dimensional Euclidean space, where each point is parameterized by Cartesian coordinates $(x, y, z)$ and may be extended with additional attributes such as RGB values, surface normals, reflectivity, etc. The data acquisition primarily relies on structured light scanning, LiDAR, or Time-of-Flight depth sensing technologies, which directly capture the spatial sampling distribution of target object surfaces. In the realm of 3D generation, Sun et al. \cite{sun2020pointgrow} proposed a point-wise generation approach to construct point clouds, enabling visualization and interpretability during the generation process. Kim et al. \cite{kim2021setvae} leveraged permutation-invariant attention modules to process unordered 3D point sets. The recent work by Nichol et al. \cite{nichol2022point} introduced a point cloud diffusion model to achieve text-to-3D point clouds generation.

\subsubsection{Voxel}

Voxels are essentially topological extensions of 2D pixels to 3D space. By dividing the Euclidean space into a uniform 3D grid array, each cubic unit, i.e. voxel, is assigned a specific attribute value (such as density, color, material properties or scalar field information) to encode the spatial occupancy state and physical properties. This structured representation exhibits strictly isotropic spatial distribution characteristics, providing an ideal input structure for volume data feature extraction based on 3D convolutional neural networks. Representative works, such as Clip-Forge \cite{sanghi2022clip}, encoded voxels into latent vectors through 3D convolutions, which are subsequently reconstructed by a decoder. The shape embeddings are then mapped to a Gaussian distribution to support the generation of diverse shapes. Wu et al. \cite{wu2016learning} employed a fully convolutional network with progressive upsampling to map latent vectors to 3D voxels. Notably, as the spatial resolution increases, the complexity of storing voxels grows at an $O(n^3)$ rate, leading to dual constraints of memory and computational power during high-precision modeling.

\subsubsection{3D Gaussian Splatting}

The 3D Gaussian ellipsoids are initialized using sparse point clouds generated by Structure-from-Motion (SfM) \cite{schonberger2016structure}. Their parametric representation encompasses attributes including center positions, covariance matrices, colors, and opacities. Through a differentiable optimization process, this approach models 3D objects as collections of anisotropic Gaussian distributions. "Splatting" refers to the process of projecting ellipsoids onto the image plane. 3D Gaussian Splatting (3DGS) \cite{kerbl20233d}, by virtue of the geometric editability enabled by explicit primitives, high-frequency detail capture capability comparable to implicit radiance fields, and real-time rendering efficiency, has been widely applied in 3D generation tasks \cite{tang2023dreamgaussian, yi2024gaussiandreamer, li2024scenedreamer360}. However, its storage complexity grows linearly with the number of Gaussian ellipsoids, and the generation quality heavily depends on point cloud priors.

\subsection{Implicit Representations}

Implicit representations implicitly encode the geometric and radiation characteristics of three-dimensional space through continuous mathematical functions $\mathcal{F}:\mathbb{R}^3\to\mathbb{R}^k$, where the output dimension $k$ corresponds to multimodal features such as signed distance, occupancy probability or radiation properties. This continuous representation paradigm naturally supports differential operations, making differentiable rendering based on volume rendering equations possible.

\subsubsection{Signed Distance Function}

The Signed Distance Function (SDF) is defined as the directed distance $\phi(\mathbf{x})$ from any point $\mathbf{x}$ to the target surface, constructing a zero-level set surface representation that satisfies $\{\mathbf{x}|\phi(\mathbf{x}) = 0\}$. The sign property of this function (negative values inside and positive values outside) not only provides a topological basis for distinguishing the interior-exterior relationship of the geometry, but its global continuity and potential differentiability also endow it with unique advantages in mathematical processing. 3D generation methods, such as SDFusion \cite{cheng2023sdfusion}, encoded SDF into latent vectors for model training, and used a decoder to recover the SDF shape representation of objects from sampled noise during inference. One-2-3-45 \cite{liu2023one} and One-2-3-45++ \cite{liu2024one} leveraged neural networks to predict SDF values on object surfaces, followed by high-fidelity mesh conversion through isosurface extraction algorithms such as Marching Cubes \cite{lorensen1998marching}.

\subsubsection{Occupancy Field}

By defining the occupancy probability function $o(\mathbf{x}) \in [0,1]$ for spatial points $\mathbf{x}$, the Occupancy Field transforms geometric surface reconstruction into a 3D binary classification problem. Mescheder et al. \cite{mescheder2019occupancy} employed deep neural networks to parameterize occupancy field, leveraging the Marching Cubes algorithm \cite{lorensen1998marching} to extract explicit meshes at the isosurface where $o(\mathbf{x}) = 0.5$.  Compared with SDF, the occupancy field replaces signed distance regression with probabilistic supervision, significantly reducing optimization complexity while demonstrating superior generalization capabilities for high-genus topologies or non-closed surfaces. However, this approach faces inherent limitations: surface localization accuracy is constrained by spatial sampling resolution, and the absence of signed distance information leads to insufficient smoothness in reconstructed surfaces.

\subsubsection{Neural Radiance Fields}

Neural Radiance Fields (NeRF) \cite{mildenhall2021nerf} has attracted widespread attention since its introduction. This method models static scenes as a mapping function $F_\Theta: (\mathbf{x}, \mathbf{d}) \mapsto (\sigma, \mathbf{c})$ that maps spatial coordinates $\mathbf{x}$ and viewing directions $\mathbf{d}$ to volume density $\sigma$ and view-dependent radiance $\mathbf{c}$, with a Multi-Layer Perceptron (MLP) serving as the function approximator. The core innovation lies in the introduction of a high-frequency positional encoding function, which enhances the MLP's geometric detail representation capabilities through Fourier feature expansion. Based on volume rendering theory, pixel color computation along a camera ray $\mathbf{r}(t) = \mathbf{o} + t\mathbf{d}$ can be discretized as:
\begin{equation}
    \begin{split}
        C(\mathbf{r}) = \sum_{i=1}^N T_i \left(1 - \exp(-\sigma_i \delta_i)\right) \mathbf{c}_i, \quad
        where \ T_i = \exp\left(-\sum_{j=1}^{i-1} \sigma_j \delta_j\right)
    \end{split}
\end{equation}
where $\delta_i$ denotes the sampling interval and $T_i$ represents the cumulative transmittance. Although NeRF achieves photorealistic novel view synthesis, its per-ray integration strategy results in significant rendering latency. Subsequent works such as Instant-NGP \cite{muller2022instant} utilized multiresolution hash encoding for accelerated training. Meanwhile, DreamFusion \cite{poole2022dreamfusion} introduced the NeRF representation into the realm of 3D generation, inspiring a series of studies.

\begin{figure}
    \centering
    \vspace{-0.3cm}
    \includegraphics[width=\textwidth]{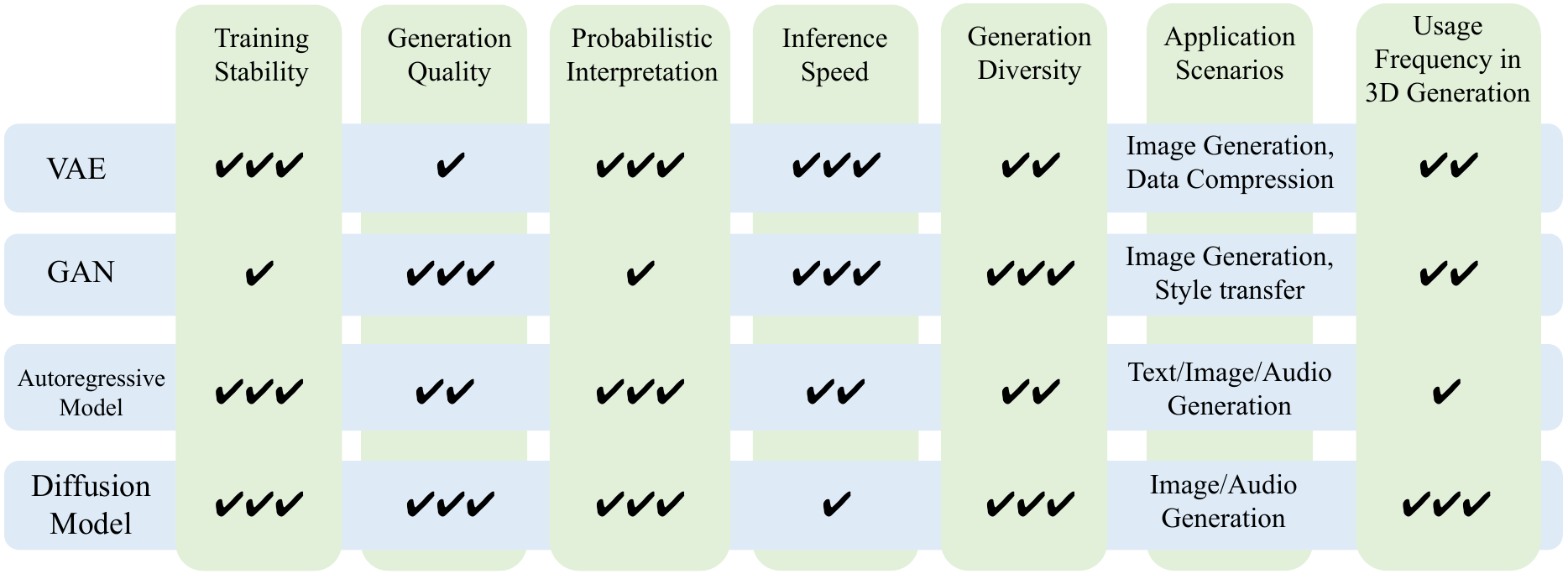}
    \vspace{-0.8cm}
    \caption{Qualitative comparison of deep generative models across different dimensions. The more symbols \protect\usym{2714} there are, the better the performance.}
    \vspace{-0.4cm}
    \label{comparison_generative_model}
\end{figure}

\subsection{Hybrid Representations}
In order to combine the advantages of explicit and implicit representations, emerging research has tended to build a hybrid representation framework. This method breaks through the inherent limitations of a single representation and dynamically combines explicit geometric primitives and implicit field functions at different spatial scales, levels of detail or functional modules to form a hierarchical scene description system.

\subsubsection{Tri-plane}

The Tri-plane representation, originally proposed by EG3D \cite{chan2022efficient}, decomposes 3D spatial features into three orthogonal Cartesian planes $(XY, XZ, YZ)$. For any 3D point $\mathbf{x}$, its corresponding feature vectors $F_{xy}$, $F_{xz}$, and $F_{yz}$ are projected onto these planes and combined via simple weighted aggregation to generate the final 3D feature vector $\mathbf{F}$. An MLP then maps $\mathbf{F}$ to point attributes such as color and density. Specifically,  EG3D \cite{chan2022efficient} innovatively integrated the tri-plane architecture with the StyleGAN2 \cite{karras2020analyzing} framework, leveraging Generative Adversarial Networks to directly synthesize tri-plane features and combining them with differentiable volume rendering for high-fidelity 3D face generation. TensoRF \cite{chen2022tensorf} employed tensor decomposition to decouple the 3D feature field into a product of planar matrices and vector modes, reducing storage complexity from $O(n^3)$ to $O(n^2)$ while maintaining reconstruction quality.

\subsubsection{DMTet}

As a groundbreaking framework for hybrid 3D representation, DMTet \cite{shen2021deep} discretizes the 3D space into structured deformable tetrahedral meshes, where each vertex is associated with implicit SDF values and their gradient information. By leveraging neural networks for collaborative optimization, it dynamically extracts explicit surface meshes through the differentiable Marching Tetrahedra algorithm \cite{doi1991efficient}, enabling efficient and high-precision geometric modeling. The key advantages of DMTet lie in its capability to handle complex 3D geometric structures and surface details, as well as its seamless integration with differentiable rendering methods. In 3D object generation approaches, both Magic3D \cite{lin2023magic3d} and Magic123 \cite{qian2023magic123} adopted a two-stage optimization framework, converting the implicit NeRF representations from the initial stage into DMTet for further refinement. The latest work Sherpa3D \cite{liu2024sherpa3d} ensured multi-view consistency of generated objects by optimizing coarse 3D object priors represented through DMTet, which were produced by the 3D diffusion model (i.e., Shap-E \cite{jun2023shap}).
\section{3D Object Generation} \label{section4}

\begin{figure}
    \centering
    \vspace{-0.3cm}
    \includegraphics[width=\textwidth]{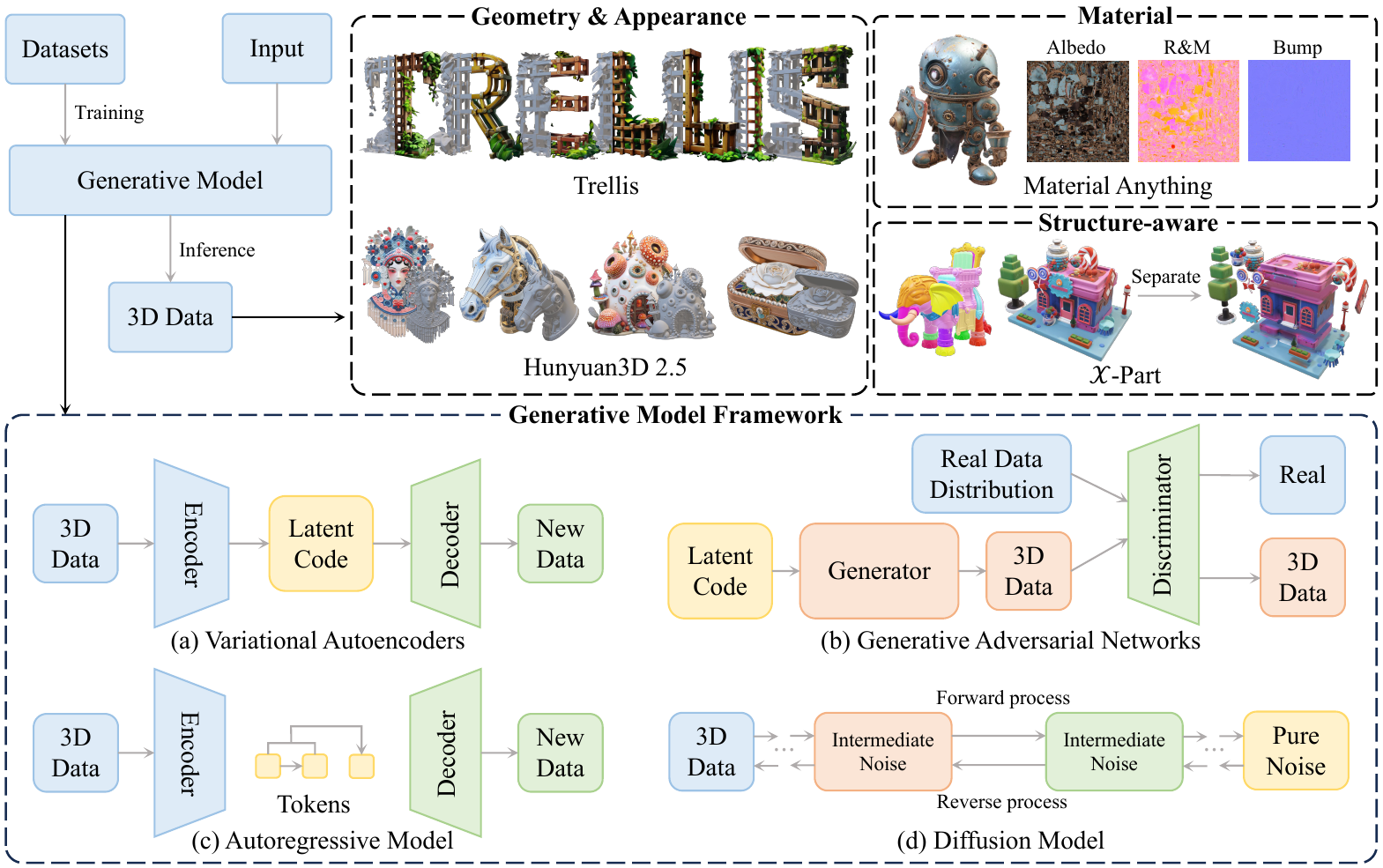}
    \vspace{-0.7cm}
    \caption{Overview of 3D object generation methods. The upper panel presents the general workflow and key application categories of data-driven generative models. The lower panel enumerates four mainstream generative model frameworks.}
    \vspace{-0.4cm}
    \label{object_generation_paradigms}
\end{figure}

In recent years, deep generative models have achieved transformative breakthroughs in 2D image generation, leveraging their sophisticated representation learning capabilities. Prominent architectures such as Variational Autoencoders, Generative Adversarial Networks, Autoregressive Models, and Diffusion Models have demonstrated unprecedented capacities in modeling intricate visual distributions. Based on the core characteristics of the aforementioned architectures, Fig. \ref{comparison_generative_model} presents a qualitative comparison of them from dimensions such as performance and probabilistic interpretability. Motivated by these advancements, an increasing number of studies have focused on integrating such data-driven generative frameworks with diverse 3D representations to expand their applicability into the domain of 3D object generation. As illustrated in Fig. \ref{object_generation_paradigms}, this section will introduce the basic theoretical framework of generative models and delve into the evolution and applications of various models across different dimensions, including geometry, joint geometry and appearance, appearance and material, and structure-aware.

\subsection{Variational Autoencoders}

Variational Autoencoders (VAEs) \cite{kingma2013auto} achieve probabilistic latent representation learning of data through an encoder-decoder architecture. The encoder maps input data to probability distribution parameters in the latent space via nonlinear transformations, with the reparameterization trick enabling sampling from this distribution to generate statistically diverse latent codes. The decoder, typically symmetric to the encoder, reconstructs latent codes into input-space data, forming an end-to-end generative system. The model optimizes the generative distribution by maximizing the evidence lower bound, thereby achieving probabilistic modeling of data distributions.

In the early exploration of 3D geometry generation, VAEs were widely employed to learn compact and continuous shape latent spaces. While VAEs exhibit significant advantages in explicit probabilistic modeling and stable training, they are confronted with limitations in high-frequency geometric details due to the smoothing effect of reconstruction losses such as Mean Squared Error. DeepSDF \cite{park2019deepsdf} innovatively proposed an Auto-Decoder framework, which directly optimized the latent codes paired with each shape to fit continuous SDF and achieved high-quality 3D object shape reconstruction via the differentiable Marching Cubes algorithm \cite{lorensen1998marching}. SetVAE \cite{kim2021setvae} further introduced hierarchical VAEs and attention mechanisms, achieving high-fidelity 3D shape generation using point cloud representations. In contrast, SDM-NET \cite{gao2019sdm} adopted a two-stage VAE architecture, which encodes part geometry and global structure separately to generate deformed meshes with semantic information.

\subsection{Generative Adversarial Networks}

Generative Adversarial Networks (GANs) \cite{goodfellow2014generative} are deep generative models based on adversarial training, which have achieved remarkable results in image generation tasks. The core idea of GANs is to construct a zero-sum game between the generator $G$ and the discriminator $D$; $G$ tries to generate samples to fool $D$, while $D$ strives to distinguish real data from generated data. This competition eventually converges to a generative Nash equilibrium, making the generated data samples close to the real distribution. The optimization goal of GAN can be formalized as the following minimax problem:
\begin{equation}
\begin{aligned}
        \min_G \max_D V(D, G) &= \mathbb{E}_{x \sim p_{\text{data}}}[\log D(x)] + \mathbb{E}_{z \sim p_z}[\log(1 - D(G(z)))]
\end{aligned}
\end{equation}
where $z$ is the latent space noise input, $p_{\text{data}}$ is the real data distribution, and $p_z$ is the noise distribution. During the training process, one side is fixed and iterates alternately. 

Building upon the groundbreaking advancements of GANs in 2D image generation, researchers have actively explored their potential in 3D geometry generation. In early studies, Wu et al. \cite{wu2016learning} pioneered the extension of 2D convolutions to 3D voxel space, utilizing 3D convolutional networks to directly generate voxelized objects, and further constructing a 3D-VAE-GAN framework to realize inference from 2D images to 3D shapes. To address the limitation of restricted resolution in voxel representation, SurfGen \cite{luo2021surfgen} employed DeepSDF \cite{park2019deepsdf} as the generator to extract 3D surface meshes. It then transformed irregular 3D meshes into regular spherical parameterizations and leveraged GCN to extract features from spherical projections. By designing adversarial loss functions based on surface geometric features, this approach directly optimized object surfaces through adversarial training, effectively addressing the insufficient geometric constraints in traditional methods.

To fully exploit 2D visual priors for guiding 3D generation, 3D-aware GANs have emerged. These methods focus on constructing implicit or explicit 3D representations through generators, followed by differentiable rendering to synthesize 2D images. The discriminator then evaluates discrepancies between generated and real images to optimize the generation process. For instance, HoloGAN \cite{nguyen2019hologan} became the first unsupervised generative model to learn 3D representations from unlabeled 2D images. It utilized 3D convolutional networks to generate 3D features from fixed tensors, applied rigid transformations to enable arbitrary pose adjustments, and finally produced images through differentiable projection. Similarly, BlockGAN \cite{nguyen2020blockgan} explicitly separated 3D objects from the scene, generated 3D features with independent generators, and combined them into a unified scene feature to achieve disentangled control. Henzler et al. \cite{henzler2019escaping} extracted 3D voxels from 2D images, incorporated differentiable rendering layers for image synthesis, and optimized 3D structures using GAN loss. To introduce semantic control, Text2Shape \cite{chen2019text2shape} further leveraged Conditional Wasserstein GAN to learn the joint embedding of text and colorful voxels, enabling the generation of 3D objects with corresponding color and shape details. With the advancement of NeRF, EG3D \cite{chan2022efficient} introduced an efficient tri-plane representation. While it can generate high-quality geometries with multi-view consistency, it suffers from bottlenecks in rendering efficiency. CLIP-NeRF \cite{wang2022clipnerf}, built upon such architectures, integrated CLIP embeddings to realize text- or image-driven manipulation of shape and appearance. To directly obtain explicit meshes, GET3D \cite{gao2022get3d} introduced a DMTet-based geometric representation coupled with tri-plane texture mapping, achieving high-fidelity 3D textured mesh generation through adversarial training. TextField3D \cite{huang2023textfield3d} further introduced a Noisy Text Field to enhance the mapping between limited 3D data and open vocabulary, and used a multi-modal discriminator to guide conditional generation based on the GET3D backbone. Addressing the demand for real-time rendering, Barthel et al. \cite{barthel2024gaussian} put forward an innovative approach. By training a sequential decoder to decode the tri-plane features of a pre-trained GAN into Gaussian attributes, this method enabled end-to-end conversion to 3DGS scenes.

In addition to the holistic generation of objects, GANs have also been demonstrated to exhibit distinct advantages in addressing the structured decomposition of objects. Early works on structure-aware generation were dedicated to learning the geometric distribution and topological relationships of parts in a low-dimensional latent space. To capture the symmetric hierarchical structure of objects, GRASS \cite{li2017grass} combined VAE with GAN, leveraging a recurrent neural network to encode object features and generating voxelized structural shapes through adversarial training. Similarly, Li et al. \cite{li2020learning} adopted a divide-and-conquer strategy, generating semantic parts individually via a VAE-GAN array and then predicting transformation parameters to assemble them into complete objects. Moreover, for point cloud representation, MRGAN \cite{gal2020mrgan} proposed a tree-structured graph convolutional GAN with multiple root nodes, realizing unsupervised disentangled generation of object parts.

\subsection{Autoregressive Models}

Autoregressive models are modeled by decomposing the joint distribution of high-dimensional data into a product of conditional distributions. Formally, given a sequence of variables $\mathbf{x} = (x_1, x_2, \ldots, x_T)$, the joint probability $p(\mathbf{x}) $ is expressed as:
\vspace{-0.2cm}
\begin{equation}
    p(\mathbf{x}) = \prod_{t=1}^T p(x_t \mid x_{<t})    
\end{equation}
\vspace{-0.2cm}

\noindent where each element $x_t$ is generated conditioned on all preceding elements  $x_{<t}$ .This sequential dependency enables autoregressive models to capture complex local and global patterns in data. 

The autoregressive framework, initially achieving remarkable success in image generation (e.g., PixelRNN \cite{van2016pixel}) and natural language processing (e.g., GPT \cite{radford2018improving}), has been progressively adapted to 3D generation tasks due to its inherent capability to model structural and sequential dependencies in 3D data. The self-attention mechanism of the Transformer \cite{vaswani2017attention}, which enables efficient long-range dependency modeling, is widely adopted as the backbone architecture for autoregressive models. In terms of point cloud generation, PointGrow \cite{sun2020pointgrow} adopted a point-wise autoregressive strategy and leveraged the self-attention mechanism to capture long-range correlations between points. For mesh data, PolyGen \cite{nash2020polygen} modeled the vertices and faces of 3D meshes as sequences respectively, and generated vertex coordinates and topological connections sequentially through two cascaded Transformers. MeshXL \cite{chen2024meshxl} proposed a Neural Coordinate Field to serialize meshes, and directly generated high-fidelity 3D meshes using a large-scale pre-trained Transformer model.

Despite the progress made in direct serial generation, compressing 3D shapes into a quantized latent space has emerged as a mainstream trend to further reduce computational complexity and improve generation resolution. AutoSDF \cite{mittal2022autosdf} and CLIP-Sculptor \cite{sanghi2022clip} employed VQ-VAE to compress SDFs or voxels into discrete latent code sequences, and then utilized Transformers to learn the prior distribution. Going a step further, CLIP-Sculptor enabled text-conditioned zero-shot generation. ShapeFormer \cite{yan2022shapeformer} proposed a sparse VQDIF representation for SDFs, which only encoded non-empty regions, and autoregressively predicted the feature sequences of missing parts via a Transformer to accomplish shape completion. ShapeCrafter \cite{fu2022shapecrafter} extended this framework to support recursive text input, where the autoregressive model gradually evolved and refined the generated shape according to incrementally added text phrases.

Autoregressive models are also frequently employed for modeling part-based structured objects. In particular, TM-NET \cite{gao2021tm} introduced a conditional autoregressive model and a quantized VAE, achieving disentangled learning of texture distribution and part geometry as well as the synthesis of high-frequency texture details. PQ-NET \cite{wu2020pq} regarded 3D objects as part sequences and utilized a Seq2Seq network to encode and decode part geometry and affine transformations. Recent works have further explored the potential of Transformers in handling complex topological sequences. BrepGPT \cite{li2025brepgpt} generated rigorous CAD boundary representations (B-rep) by autoregressively predicting half-edges, while Wang et al. \cite{wang2025autoregressive} leveraged an hourglass-shaped Transformer to model the hierarchical branch sequences and dynamic growth processes of trees.

Furthermore, to address the limitations of learning-based methods in terms of physical properties and topological editability, some works have shifted toward symbolic generation and methods leveraging large models for logical reasoning. MeshCoder \cite{dai2025meshcoder} explored inverse graphics, utilizing LLMs to convert point clouds into Blender Python code \cite{blender2018}, enabling editable geometric reconstruction. Taking advantage of the reasoning capabilities of Vision Language Models (VLMs), Articulate-Anything \cite{le2024articulate} retrieved or generated components by writing Python code and assembled them into URDF files, iteratively refining joint parameters with simulation feedback; Articulate AnyMesh \cite{qiu2025articulate} directly inferred joint parameters and component segmentation from input meshes, achieving articulated object modeling for open-set categories. For the stringent requirements of physical realism in robotic simulation, Infinigen-Sim \cite{joshi2025infinigen} and Infinigen Mobility \cite{lian2025infinite} adopted rule-based procedural generation pipelines, which generate articulated objects with precise kinematic trees, physical properties, and photorealistic textures via parameterized rules, supporting the construction of embodied intelligence datasets.

\subsection{Diffusion Models}\label{DM}

Most diffusion models currently adopted are based on DDPM \cite{ho2020denoising}, whose theoretical framework traces back to the diffusion probabilistic model proposed in \cite{sohl2015deep}. Diffusion models consist of two processes: the forward process and the reverse process. In the forward diffusion process, data $x_0$ is gradually transformed into pure noise $x_T$ via a fixed Markov chain, with Gaussian noise of increasing variance added at each step:
\vspace{-0.2cm}
\begin{equation}
    \begin{aligned}
        q(x_t | x_{t-1}) &= \mathcal{N}(x_t; \sqrt{1-\beta_t}x_{t-1}, \beta_t \mathbf{I}) \\
        q(x_{1:T} | x_{0}) &= \prod_{t=1}^T q(x_t | x_{t-1})
    \end{aligned}
\end{equation}
\vspace{-0.2cm}

\noindent where $\beta_t$ denotes the variance at timestep $t$, with $0 < \beta_1 < \cdots < \beta_T < 1$ controlling the noise schedule. The reverse process aims to denoise the data, where the diffusion model learns a neural network $p_{\theta}(x_{t-1}|x_t)$ to approximate the true reverse transition $q(x_{t-1}|x_t)$:  
\begin{equation}
\begin{aligned}
    p_{\theta}(x_{t-1}|x_t) &=\mathcal{N}(x_{t-1}; \mu_{\theta}(x_t,t),\Sigma_{\theta}(x_t,t))\\
    p_\theta(x_{0:T}) &= p(x_T) \prod_{t=1}^T p_\theta(x_{t-1} | x_t)
\end{aligned}
\end{equation}
with $p(x_T)=\mathcal{N}(x_T;0,\mathbf{I})$. The model is optimized by minimizing the variational lower bound, which is simplified to the MSE of the predicted noise $\epsilon$:
\vspace{-0.2cm}
\begin{equation}
\begin{aligned}
    \mathcal{L}_{\text{simple}} = \mathbb{E}_{t,x_0,\epsilon}\left[ \| \epsilon - \epsilon_\theta(x_t,t) \|^2 \right]
\end{aligned}
\end{equation}

\subsubsection{Geometry-only}
\ 
\newline
The generation of 3D geometric structures serves as the foundation of object generation, and its core lies in how to utilize compact and efficient mathematical representations to model the distribution of object shapes. Zhou et al. \cite{zhou20213d} pioneered a 3D shape generation method that integrates Point-Voxel representation with diffusion models. By enabling efficient conversion between point cloud and voxel features, it unified unconditional generation and point cloud completion tasks. To further handle high-resolution geometries and reduce computational costs, subsequent works have widely adopted the Latent Diffusion strategy, which first compresses 3D data into a low-dimensional latent space before conducting diffusion training. For instance, SDFusion \cite{cheng2023sdfusion} leveraged VQ-VAE to compress 3D shapes into compact latent representations, achieving high-fidelity generation under multi-modal conditions (text, images, and partial shapes). Michelangelo \cite{zhao2023michelangelo} further proposed the Shape-Image-Text Aligned VAE and Aligned Shape Latent Diffusion Model. By constructing a semantically aligned latent space, it effectively bridged cross-modal discrepancies.

To further enhance generation efficiency and optimize sampling trajectories, state-of-the-art works such as \cite{ye2025hi3dgen, wu2025direct3d, chen2025ultra3d} have shifted toward Flow Matching \cite{lipman2022flow} or Rectified Flow \cite{liu2022flow}. Unlike DDPM, Flow Matching constructs a continuous mapping from the noise distribution to the data distribution by learning a time-dependent velocity field \(v_\theta\), with its training objective typically regressing to $x_0-\epsilon$:
\begin{equation}
\begin{aligned}
    \mathcal{L}_{FM} = \mathbb{E}_{t, x_0,\epsilon} [\| v_\theta(x_t,t) - (x_0-\epsilon) \|^2]
\end{aligned}
\end{equation}
This allows the generation process to follow a straighter trajectory, thereby significantly reducing the number of sampling steps. Combining sparse voxel representation with flow matching models, Hi3DGen \cite{ye2025hi3dgen} proposed a normal bridging strategy, utilizing a normal-regularized latent flow matching model to generate detail-rich, high-fidelity geometries in the sparse voxel space. Direct3D-S2 \cite{wu2025direct3d} introduced a Spatial Sparse Attention mechanism, while Ultra3D \cite{chen2025ultra3d} incorporated a Part Attention mechanism. Both optimize the attention computation in Diffusion Transformer (DiT), substantially lowering the computational complexity when processing large-scale sparse voxels, and greatly improving efficiency while maintaining generation quality.

\subsubsection{Joint Geometry and Appearance}
\
\newline
\textbf{SDS-based Optimization.} Methods based on other generative architectures are primarily limited to modeling object geometry and often suffer from a lack of texture detail. In contrast, approaches elevate the powerful 2D image generation priors of diffusion models to 3D domain not only enable high-precision geometric reconstruction of objects but also produce textures with rich details. DreamFusion \cite{poole2022dreamfusion} pioneered the use of pre-trained 2D diffusion models to guide text-to-3D object generation, introducing the groundbreaking concept of Score Distillation Sampling (SDS). Concretely, consider a differentiable 3D representation parameterized by $\theta$ and a rendering function $g$, the rendered image can be expressed as $x = g(\theta, c)$ for a given camera pose $c$. SDS optimizes the 3D representation parameters $\theta$ using a fixed-parameter 2D diffusion model $\phi$. The core idea lies in adjusting $\theta$ by computing specific gradients to align the rendered image $x$ with the distribution of the 2D text-to-image model. The specific gradient calculation formula is defined as:
\vspace{-0.2cm}
\begin{equation}
\begin{aligned}
    \nabla_{\theta} \mathcal{L}_{SDS}(\phi,x=g(\theta,c))=\mathbb{E}_{t, \epsilon, c}\left[w(t)\left(\hat\epsilon_{\phi}\left(x_{t} ; y, t\right)-\epsilon\right) \frac{\partial x}{\partial \theta}\right]
\end{aligned}
\end{equation}
\vspace{-0.3cm}

\noindent where $\mathbb{E}_{t, \epsilon, c}$ denotes the expectation over timestep $t$, noise $\epsilon$, and camera pose $c$; $w(t)$ is a time-dependent weighting function that adjusts the importance of information at different timesteps; $\hat\epsilon_{\phi}\left(x_{t} ; y, t\right)$ represents the noise predicted by the 2D diffusion model $\phi$ at timestep $t$ under text prompt $y$; $\epsilon$ is the ground-truth noise added to the rendered image $x$; and $\frac{\partial x}{\partial \theta}$ is the gradient of the rendering function $g$ with respect to parameters $\theta$, quantifying the impact of parameter changes on the rendered image. Through iteratively computing gradients and updating $\theta$, the rendered image  progressively aligns with the distribution of images generated by the 2D text-to-image model, thereby optimizing the 3D representation.

This groundbreaking approach laid a theoretical foundation for 3D generation based on pre-trained models. Numerous follow-up studies \cite{melas2023realfusion, tang2023make, chen2023fantasia3d, qian2023magic123, shi2023mvdream, tang2023dreamgaussian, yi2024gaussiandreamer, sun2023dreamcraft3d, lin2023magic3d, liu2024unidream, liu2024sherpa3d} have conducted diversified explorations on the selection of 3D representations to improve generation quality and efficiency. Similar to DreamFusion, RealFusion \cite{melas2023realfusion} optimized NeRF to represent objects, achieving full 360° reconstruction of objects from a single image. Make-it-3D \cite{tang2023make} leveraged diffusion priors, reference view supervision, and depth priors to optimize NeRF for obtaining a coarse model, which was then converted into a textured point cloud for further refinement. Compared to NeRF-based approaches, methods employing 3DGS for object modeling exhibit significantly faster convergence. DreamGaussian \cite{tang2023dreamgaussian} introduced an efficient algorithm to convert generated Gaussians into textured meshes. GaussianDreamer \cite{yi2024gaussiandreamer} utilized 3D diffusion models (i.e., Shap-E \cite{jun2023shap}) to generate coarse 3D instances, converted them into point clouds, enhanced the point clouds via noisy point growth and color perturbation for 3D Gaussians initialization, and further optimized the 3D Gaussians using SDS. This pipeline achieves real-time rendering and generates 3D instances on a single GPU within 15 minutes. Several works employed hybrid 3D representations, such as DMTet, for object modeling. Specifically, Fantasia3D \cite{chen2023fantasia3d} decoupled object geometry and texture. For geometry, it encoded extracted surface normals into the input of an image diffusion model; for texture, it introduced spatially varying BRDFs to learn surface materials for physics-based rendering. Similarly, DreamCraft3D \cite{sun2023dreamcraft3d} decomposed object generation into two stages: geometric sculpting and texture boosting. To address the low-resolution output issue of DreamFusion, Magic3D \cite{lin2023magic3d} adopted a two-stage optimization framework: first obtaining a coarse model via low-resolution diffusion priors and a hash-grid-encoded neural field, then refining it into a high-resolution textured 3D mesh using latent diffusion models (such as Stable Diffusion \cite{rombach2022high}). Magic123 \cite{qian2023magic123} also adopted a two-stage strategy, combining 2D and 3D diffusion priors to generate 3D content from a single pose-free image. In addition, to address the problem of geometric consistency during the generation process, Sherpa3D \cite{liu2024sherpa3d}, DreamControl \cite{huang2024dreamcontrol}, and Dream3D \cite{xu2023dream3d} all emphasized the importance of geometric priors. Concretely, Sherpa3D leveraged rough 3D models to provide geometric guidance; DreamControl resolved the multi-faced Janus problem via coarse-grained 3D priors; and Dream3D combined stylized views generated by Stable Diffusion \cite{rombach2022high} with CLIP guidance, which significantly improved the geometric accuracy of the generated content under zero-shot conditions.

\begin{table}
\centering
\vspace{-0.2cm}
\caption{The gradient calculation formula of the state-of-the-art score distillation methods. Additionally, using Instant-NGP \cite{muller2022instant} as the 3D representation, we conduct a quantitative comparison of different score distillation methods under specific prompts. The evaluation metrics include "Sim", which indicates the semantic similarity between generated images and the text, and "R@1" representing the CLIP recall rate. This metric measures the classification accuracy of predicting the correct text prompt by applying the CLIP model to rendered images.}
\vspace{-0.2cm}
\centering
\resizebox{\textwidth}{!}{
\scriptsize
\begin{tabular}{@{}l l >{\raggedright}m{4.5cm} c c@{}}
\toprule
\textbf{Methods} & \textbf{Gradient Calculation Formula} & \textbf{Notation Explanation} & \textbf{Sim} $\uparrow$ & \textbf{R@1} $\uparrow$ \\
\midrule
SDS \cite{poole2022dreamfusion} & 
$\mathbb{E}_{t, \epsilon, c}\left[w(t)\left(\hat\epsilon_{\phi}\left(x_{t} ; y, t\right)-\epsilon\right) \frac{\partial x}{\partial \theta}\right]$ & 
Section \ref{DM} & 0.288 & 1.000 \\
\midrule 
CSD \cite{yu2023text} & 
$\mathbb{E}_{t, \epsilon, c}\left[w(t)\left(\hat\epsilon_{\phi}\left(x_{t} ; y, t\right)-\hat\epsilon_{\phi}\left(x_{t} ; \emptyset, t\right)\right) \frac{\partial x}{\partial \theta}\right]$ & 
$\hat{\epsilon}_{\phi}(x_{t}; \emptyset, t)$: The noise estimate at timestep $t$ without a text prompt $y$. & 0.280 & 0.936 \\
\midrule
ISM \cite{liang2024luciddreamer} & 
$\mathbb{E}_{t, \epsilon, c}\left[w(t)\left(\hat\epsilon_{\phi}\left(x_{t} ; y, t\right)-\hat\epsilon_{\phi}\left(x_{s} ; \emptyset, s\right)\right) \frac{\partial x}{\partial \theta}\right]$ & 
$\hat{\epsilon}_{\phi}(x_{s}; \emptyset, s)$: The noise estimate at timestep $s$ without a text prompt $y$. & - & - \\
\midrule 
VSD \cite{wang2023prolificdreamer} & 
$\mathbb{E}_{t, \epsilon, c}\left[w(t)\left(\hat\epsilon_{\phi}\left(x_{t} ; y, t\right)-\epsilon_{\phi'}\left(x_{t} ; y, t,c\right)\right) \frac{\partial x}{\partial \theta}\right]$ & 
$\epsilon_{\phi'}(x_{t}; y, t, c)$: The noise estimate predicted by a LoRA-fine-tuned model. & 0.276 & 0.932 \\
\midrule
ASD \cite{ma2024scaledreamer} & 
$\mathbb{E}_{t, \epsilon, c}\left[w(t)\left(\hat\epsilon_{\phi}\left(x_{t} ; y, t\right)-\hat\epsilon_{\phi}\left(x_{t+\Delta t} ; y, t+\Delta t\right)\right) \frac{\partial x}{\partial \theta}\right]$ & 
$\hat{\epsilon}_{\phi}(x_{t+\Delta t}; y, t+\Delta t)$: The noise estimate at a timestep $t+\Delta t$ with a text prompt $y$. & 0.289 & 1.000 \\
\bottomrule
\end{tabular}}
\vspace{-0.6cm}
\label{sds_variant}
\end{table}

\textbf{Improved Variants of SDS.} Although SDS-based 3D generation methods demonstrate significant advantages in geometric representation optimization, their inherent defect of pseudo-ground-truth (pseudo-GT) distribution inconsistency leads to degraded quality and over-smoothing in model outputs. A series of representative works have proposed variants to tackle these shortcomings, as summarized in Table \ref{sds_variant}. Yu et al. \cite{yu2023text} revealed that the effectiveness of SDS stems from its two gradient components: the generative prior $\hat\epsilon_{\phi}(x_{t} ; y, t) - \epsilon$ and the classifier score $\hat\epsilon_{\phi}(x_{t} ; y, t) - \hat\epsilon_{\phi}(x_{t} ; \emptyset, t)$. Experiments demonstrate that SDS heavily depends on classifier-free guidance, causing the classifier score to dominate the optimization direction while the generative prior contributes minimally. Consequently, Classifier Score Distillation (CSD) discarded the generative prior and solely used the classifier score for optimization. LucidDreamer \cite{liang2024luciddreamer} introduced Interval Score Matching (ISM) to improve SDS. This method first employs DDIM inversion to generate reversible diffusion trajectories, reducing the averaging effect caused by pseudo-GT inconsistency. Additionally, ISM matches between two interval steps along the diffusion trajectory, avoiding high reconstruction errors from single-step optimization. As an extension, GaussianDreamerPro \cite{yi2024gaussiandreamerpro} addressed the blurry edges in GaussianDreamer \cite{yi2024gaussiandreamer} by binding Gaussians to the surface of base 3D assets. By incorporating geometric constraints and optimizing Gaussian ellipsoids via ISM, it generated enhanced 3D objects. SDS also faces challenges with over-saturation and low diversity. To overcome these limitations, ProlificDreamer \cite{wang2023prolificdreamer} proposed Variational Score Distillation (VSD), which modeled 3D parameters $\theta$ as distributions and optimized their distribution via particle variational inference. Combined with Low-Rank Adaptation (LoRA) \cite{hu2022lora, ryu2023low} driven score estimation, VSD achieves more precise optimization directions. Sculpt3D \cite{chen2024sculpt3d} generated detail-rich 3D assets via retrieval-augmented strategies and VSD optimization. The latest work, ScaleDreamer \cite{ma2024scaledreamer}, pointed out that VSD’s reliance on LoRA fine-tuning compromises the generalization ability of pre-trained models to diverse text prompts, leading to training instability and mode collapse. Asynchronous Score Distillation (ASD) aims to resolve these issues by exploiting the observation that diffusion models exhibit lower noise prediction errors at earlier timesteps. By shifting the current timestep $t$ forward to $t + \Delta t$, ASD reduces noise prediction errors without altering the pre-trained model weights, thus maintaining the model's capabilities while generating high-quality and diverse 3D content.

\textbf{Multi-view Consistency.} The text-to-3D generation task faces significant modal mismatch challenges, rooted in the inherent contradiction between the sparsity of textual semantic guidance and the high-dimensional complexity of 3D geometric space. In contrast, the image-to-3D generation paradigms effectively enhance controllability by introducing strong visual priors. However, due to the lack of multi-view semantic associations and insufficient 3D geometric awareness, 2D lifting approaches often encounter the multi-faced Janus problem. To address this, numerous studies have attempted to fine-tune pre-trained diffusion models to generate multi-view consistent images from single images. MVDream \cite{shi2023mvdream} constructed a diffusion model capable of generating multi-view images conditioned on text prompts. Zero-1-2-3 \cite{liu2023zero} equipped Stable Diffusion models with camera viewpoint control, enabling novel view synthesis from a single input image and specified camera transformations, which can be applied to 3D reconstruction tasks. Its successor, Zero123++ \cite{shi2023zero123++}, refined the Stable Diffusion 2 $v$-model through multi-view layout generation, noise schedule adjustment, scaled reference attention, and global conditioning mechanisms to produce more compliant multi-view images. One-2-3-45 \cite{liu2023one} and its derivative One-2-3-45++ \cite{liu2024one} adopted Zero-1-2-3 as a multi-view diffusion model to generate 3D textured meshes from single images. GeoDream \cite{ma2023geodream} integrated the aforementioned works. Concretely, it first extracted 3D geometric priors via One-2-3-45 \cite{liu2023one}, followed by optimization with multi-view diffusion models \cite{shi2023mvdream, liu2023zero, shi2023zero123++} combined with the VSD loss \cite{wang2023prolificdreamer}. SyncDreamer \cite{liu2023syncdreamer} generated multiple images simultaneously during reverse diffusion by constructing shared noise predictors, synchronizing noise predictions via 3D-aware attention mechanism to ensure multi-view consistency. Wonder3D \cite{long2024wonder3d} trained a cross-domain diffusion model to generate multi-view consistent normal maps and color maps, which were fused through SDF optimization to reconstruct 3D object geometries. To improve lighting realism, UniDream \cite{liu2024unidream} built an albedo-normal aligned multi-view diffusion model, combining a Transformer-based reconstruction model with SDS optimization to endow objects with Physically-Based Rendering (PBR) materials using Stable Diffusion. DMV3D \cite{xu2023dmv3d} generated tri-plane representations end-to-end from text or single-image inputs. MVDiffusion++ \cite{tang2024mvdiffusion++} focused on high-resolution dense view synthesis through a pose-free multi-view diffusion model and view dropout strategy. Despite these advances, existing methods like SyncDreamer \cite{liu2023syncdreamer} and Wonder3D \cite{long2024wonder3d} still produced coarse geometry and low-resolution textures due to residual local inconsistencies and architectural limitations. In contrast, Unique3D \cite{wu2024unique3d} generated multi-view images and normal maps via a multi-view diffusion model, enhanced resolution through multi-level upsampling, and reconstructed 3D meshes from high-resolution data using the instant and consistent mesh reconstruction algorithm. Fancy123 \cite{yu2025fancy123} eliminated multi-view ghosting artifacts via 2D deformation, resolving local inconsistencies.

\textbf{Feed-forward.} Compared with optimization-based methods, feed-forward generation models offer remarkable advantages in inference speed. Early works such as Point-E \cite{nichol2022point} and Shap-E \cite{jun2023shap} explored native 3D diffusion models, which directly generate outputs in point cloud or implicit parameter spaces with ultra-fast speed but limited geometric and texture details. Recently, breakthroughs have been achieved in Transformer-based Large Reconstruction Models. InstantMesh \cite{xu2024instantmesh} could directly regress Tri-plane representations from image features; 3DTOPIA-XL \cite{chen20253dtopia} enabled high-resolution generation based on PrimX primitive representation; and Hunyuan3D 2.5 \cite{lai2025hunyuan3d} proposed a shape foundation model LATTICE, which achieves industry-grade generation quality when combined with multi-view texture synthesis. Moreover, the generation paradigm based on Flow Matching has demonstrated enormous potential. TripoSG \cite{li2025triposg} leveraged large-scale datasets to train a Rectified Flow model, realizing high-fidelity SDF generation. Trellis \cite{xiang2025structured} defined a unified structured latent representation for 3D assets, and employed a Flow Matching Transformer optimized for sparse data to generate structures and features separately, supporting decoding into multiple formats (i.e., 3DGS/Radiance Field/Mesh). UniLat3D \cite{wu2025unilat3d} further proposed a unified latent space representation for both geometry and appearance, which is directly generated via a single-stage Flow Matching model, thus completely addressing the alignment issues inherent in cascaded generation pipelines.

\subsubsection{Appearance and Material}
\
\newline
In addition to the research paradigm of jointly generating object geometry and appearance, several studies have focused on texture enhancement for given 3D geometric structures. This field primarily leverages pre-trained 2D diffusion models as texture priors and imparts high-quality appearances to base meshes via texture mapping techniques, with core challenges lying in addressing texture consistency and seam artifacts across multiple viewing perspectives.

Early explorations mainly adopted an iterative "project-paint-backproject" strategy. A representative work TEXTure \cite{richardson2023texture}, employed a depth-aware diffusion model to generate textures through view-wise iterative refinement. To mitigate inconsistencies and cumulative errors caused by fixed viewpoint sequences, Text2Tex \cite{chen2023text2tex} introduced an automatic viewpoint selection and dynamic mask generation mechanism. InTeX \cite{tang2024intex} constructed a unified depth-aware framework to support interactive texture editing, while Make-A-Texture \cite{xiang2025make} compressed the generation latency to the second level by optimizing viewpoint sequences and adopting fast backprojection techniques. For specialized applications, Ge et al. \cite{ge2024creating} aligned images with 3D models via CLIP \cite{radford2021learning}, and combined sticker generation with UV printing to realize the creation of Lego minifigures from a single input image.

However, iterative generation methods often struggle to ensure global consistency. To address this issue, research focus has gradually shifted toward synchronized multi-view generation. TexFusion \cite{cao2023texfusion} and SyncMVD \cite{liu2024text} proposed fusing latent features of overlapping regions during the denoising process of diffusion models to enforce structural consensus across different views. Building on this foundation, MVPaint \cite{cheng2025mvpaint} further introduced a synchronized multi-view generation module, and combined it with spatially-aware 3D inpainting and UV refinement algorithms, effectively resolving texture loss and seam artifacts in unobserved regions. TexGen \cite{huo2024texgen} broadcasted appearance information via an attention-guided multi-view sampling strategy and proposed a unique texture-aware noise resampling technique to preserve rich high-frequency details. In addition, Paint3D \cite{zeng2024paint3d} and Meta 3D TextureGen \cite{bensadoun2024meta} adopted a two-stage coarse-to-fine framework: first generating consistent coarse-grained textures, then leveraging a specially trained UV-space diffusion model for inpainting and super-resolution. At the architectural level, MV-Adapter \cite{huang2025mv} presented a plug-and-play adapter that can be applied to text- or image-guided 3D texture generation. RomanTex \cite{feng2025romantex} incorporated 3D-aware rotational positional encoding and a decoupled attention mechanism to explicitly inject geometric consistency at the feature level.

\begin{figure}
    \centering
    \vspace{-0.3cm}
    \includegraphics[width=\textwidth]{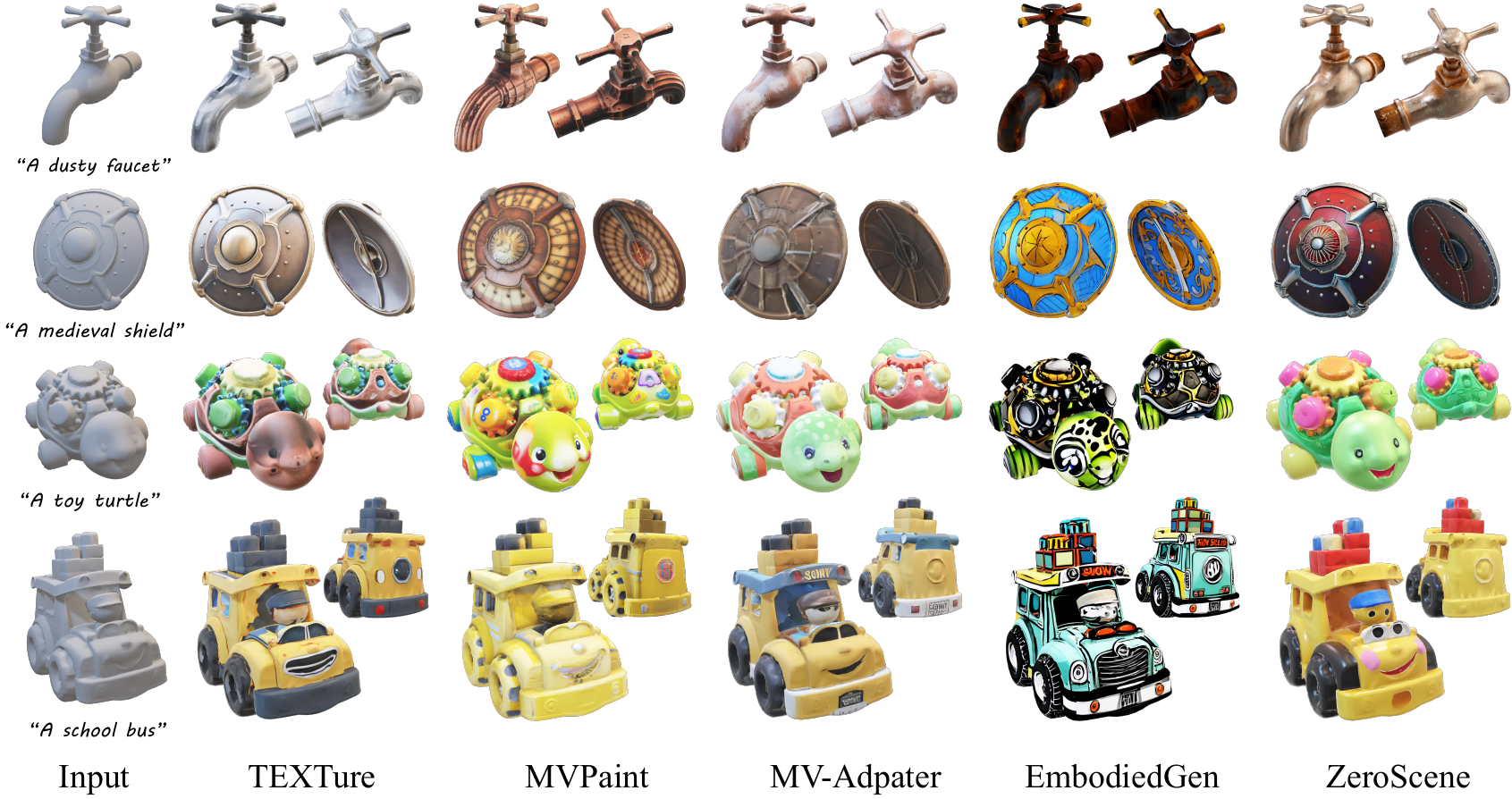}
    \vspace{-0.7cm}
    \caption{Qualitative comparison of generation methods for object appearance and material.}
    \vspace{-0.5cm}
    \label{texture_qualitative_comparison}
\end{figure}

With the escalating demands of rendering, generating only RGB color appearance can no longer meet industrial standards. Recent works have focused on producing high-fidelity PBR materials that include attributes such as albedo, roughness, and metallicity. Paint-it \cite{youwang2024paint} optimized texture maps of deep convolutional PBR using Score SDS loss, effectively filtering noise and generating physics-compliant materials. TexGaussian \cite{xiong2024texgaussian} integrated lighting attributes into the parameters of traditional 3DGS to support PBR. MaPa \cite{zhang2024mapa} achieved highly editable structured material generation by producing procedural material maps. Material Anything \cite{huang2024material} and MaterialMVP \cite{he2025materialmvp} realized material recovery and generation for objects under arbitrary lighting conditions through the introduction of confidence masks and dual-channel multi-view diffusion models, respectively. In contrast to the aforementioned projection-based approaches, Mitchel et al. \cite{mitchel2024single} proposed the concept of field latent and directly constructed an intrinsic diffusion model on the tangent vector field of mesh surfaces, enabling fully intrinsic texture generation. To enable readers to more intuitively grasp the capabilities of diffusion-based appearance and material painting methods, we present qualitative comparisons of several representative works in Fig. \ref{texture_qualitative_comparison}.

\subsubsection{Structure-aware}
\
\newline
With the advancement of diffusion models, research focus has shifted toward leveraging their powerful distribution modeling capabilities to generate structured objects with complex articulations. Methods in this category can be subdivided into three sub-directions based on technical routes. First is latent space and hybrid diffusion generation. Huang et al. \cite{huang2025part} proposed a 3D latent diffusion model based on neural voxel fields and designed a part-aware decoder to guide high-resolution generation; AutoPartGen \cite{chen2025autopartgen} and PartDiffuser \cite{yang2025partdiffuser} adopted a hybrid strategy combining autoregressive and diffusion models. The former uses autoregressive models to plan layouts and diffusion model to generate meshes, while the latter discretizes meshes into tokens and balances topology and details through inter-part autoregressive and intra-part parallel diffusion. For the assembly task of rigid parts, Assembler \cite{zhao2025assembler} employed DiT to predict the Euclidean distribution of sparse anchor point clouds and efficiently recovered part poses via least squares, enabling large-scale and generalizable object assembly. Second is structure reconstruction based on image/video priors. PartGen \cite{chen2025partgen} and Part123 \cite{liu2024part123} utilized multi-view diffusion models to generate consistent segmentation maps, reconstructing 3D parts through full-modal completion or contrastive learning, respectively; CADDreamer \cite{li2025caddreamer} focused on the engineering design domain, using multi-view diffusion models to generate normal and semantic maps and reconstructing compact CAD B-rep with geometric optimization. PartComposer \cite{liu2025partcomposer} addressed the scarcity of single-image data by learning disentangled part concepts through mutual information maximization and generating structurally reasonable 2D composite images, providing high-quality visual priors for downstream 3D modeling. SPARK \cite{he2025spark} guided geometric generation by integrating structure graphs and part images extracted by VLMs, and further introduced differentiable rendering to optimize joint parameters for ensuring motion consistency. DreamArt \cite{lu2025dreamart} incorporated a video diffusion model to predict object motion, thereby inferring articulation structures and generating textured meshes. The third and most cutting-edge direction is the Flow Matching-based generation, encompassing works such as HoloPart \cite{yang2025holopart}, Tang et al. \cite{tang2025efficient}, PartCrafter \cite{lin2025partcrafter}, OmniPart \cite{yang2025omnipart}, X-Part \cite{yan2025x}, and FullPart \cite{ding2025fullpart}. These methods are typically built on the DiT architecture and achieve efficient generation of high-quality structured objects from single images or text through mechanisms such as dual volume packing \cite{tang2025efficient}, local-global attention \cite{lin2025partcrafter}, semantic feature injection with bounding boxes \cite{yang2025omnipart, yan2025x}, center-corner encoding \cite{ding2025fullpart}, etc. Notably, HoloPart \cite{yang2025holopart} specifically focused on the amodal task of completing full geometry from partial observations.
\section{3D Scene Generation} \label{section5}

\begin{figure}
    \centering
    \vspace{-0.2cm}
    \includegraphics[width=\textwidth]{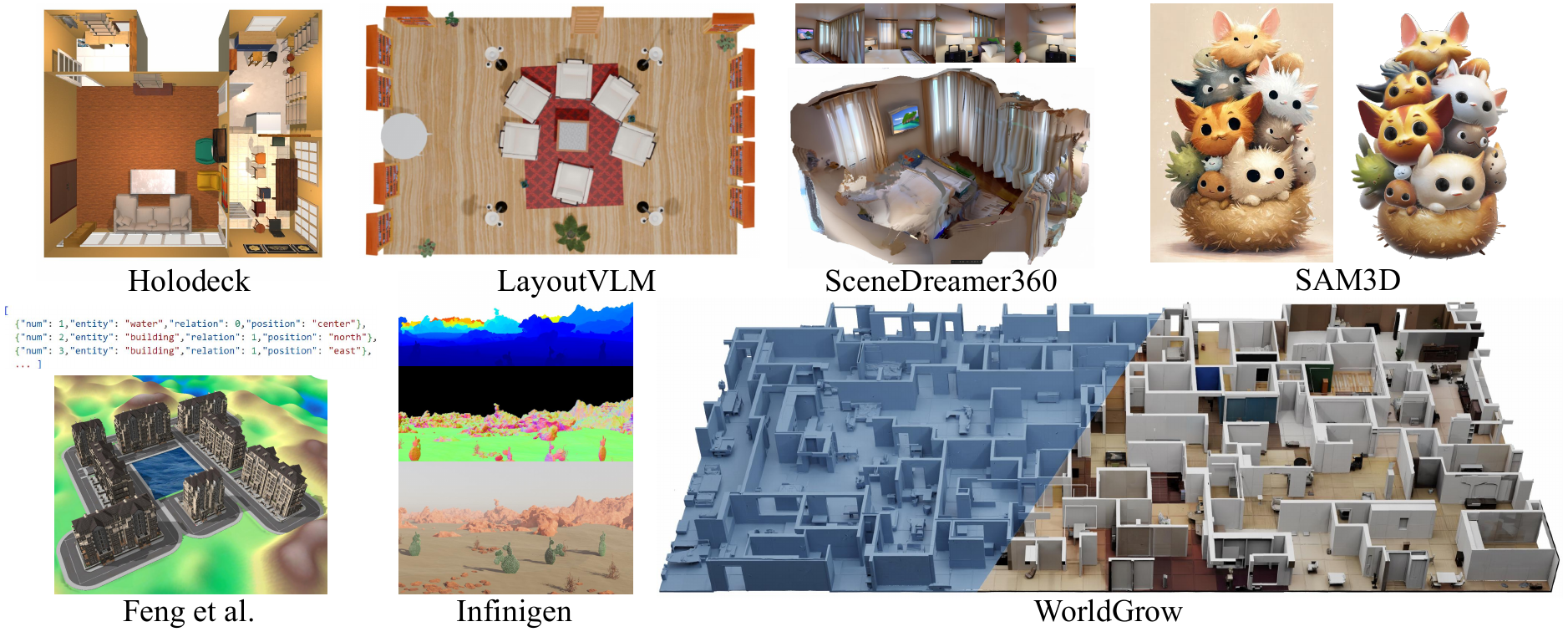}
    \vspace{-0.7cm}
    \caption{3D scene generation methods. Layout-guided: Holodeck \cite{yang2024holodeck}, LayoutVLM \cite{sun2025layoutvlm}; 2D prior-based: SceneDreamer360 \cite{li2024scenedreamer360}, SAM 3D \cite{chen2025sam}; Rule-driven: Feng et al. \cite{feng2025text}, Infinigen \cite{raistrick2023infinite}. Moreover, WorldGrow \cite{li2025worldgrow} can generate infinite scenes.}
    \vspace{-0.6cm}
    \label{scene_pre}
\end{figure}

\begin{wrapfigure}{r}{9.7cm}
    \centering
    \vspace{-0.1cm}
    \includegraphics[width=0.7\textwidth]{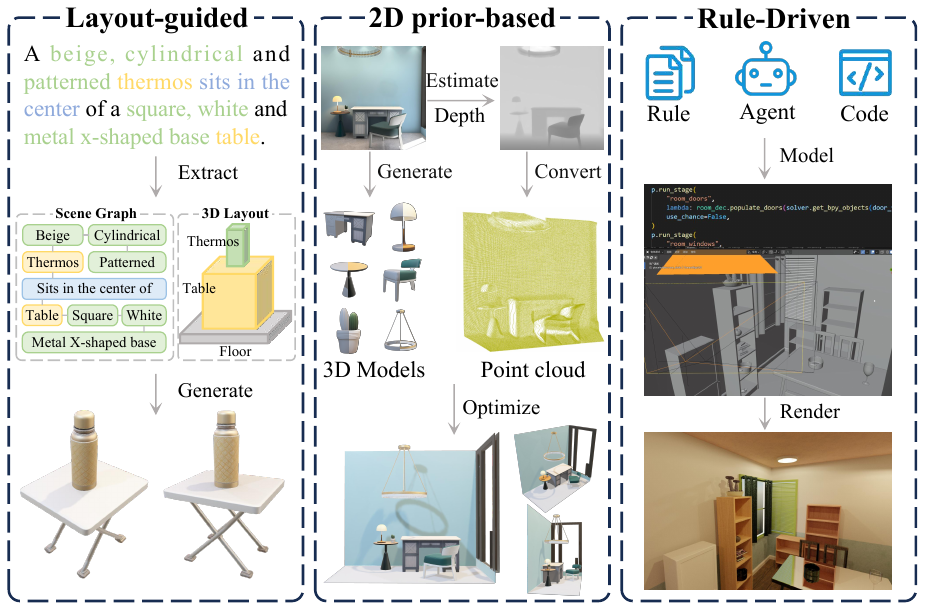}
    \vspace{-0.8cm}
    \caption{Three major paradigms of 3D scene generation methods.}
    \vspace{-0.5cm}
    \label{scene_generation_paradigms}
\end{wrapfigure}

The key to 3D scene generation lies in the collaborative construction of objects and spatial environments. Fig. \ref{scene_pre} presents the results produced by various 3D scene generation methods. From the perspective of generation mechanisms, existing methods can be categorized into three primary routes. The first is layout-guided generation, a paradigm that prioritizes the construction of a structured intermediate representation of the scene, which acts as a strong geometric and semantic constraint to guide the generation of objects or the filling of scene content. The second is scene generation based on 2D priors, the core of which is to extract the abundant visual knowledge contained in images or diffusion models and lift 2D representations to 3D space. Distinct from these two technologies that rely on data-driven generative models, rule-driven modeling achieves controllable generation of complex scene content through predefined finite rule sets. This section will systematically elaborate on these three categories of 3D scene generation paradigms, as shown in Fig. \ref{scene_generation_paradigms}.

\subsection{Layout-guided Generation}

Layout-guided generation can be defined as a technical paradigm for constructing complete scenes through explicit structured representations. Such methods typically utilize either predefined scene layouts or those learned from model-extracted scene features to model the semantic correlations and spatial topological relationships between objects. 


As an early exploration in indoor scene synthesis, Wang et al. \cite{wang2018deep} adopted autoregressive decision-making, leveraged CNNs to extract features of the orthogonal top-down view of scenes, and sequentially accomplished key operations such as object category selection, instance retrieval, orientation adjustment, and collision detection, thereby establishing the prototype of iterative layout generation. To capture more complex spatial distributions, GRAINS \cite{li2019grains} and ATISS \cite{paschalidou2021atiss} employed recurrent VAE and autoregressive Transformers respectively to learn the generative distribution from room-level structures to object attributes, becoming foundational works in this field. Also based on the Transformer architecture, LEGO-Net \cite{wei2023lego} learned to rearrange cluttered room layouts into neat states that comply with human aesthetic rules; RoomDesigner \cite{zhao2024roomdesigner} adopted lightweight anchor-latent variables to represent furniture attributes; and Forest2Seq \cite{sun2024forest2seq} proposed parsing scenes into forest structures and capturing hierarchical dependencies between objects through sequential generation. In recent years, research utilizing scene layouts as spatial geometric priors has shown a diversified development trend. PhyScene \cite{yang2024physcene} seamlessly integrated collision avoidance, room layout, and accessibility constraints into the diffusion process, exploring physical plausibility and interactivity in 3D indoor scene synthesis while providing high-quality training data for embodied AI. DiffuScene \cite{tang2024diffuscene} innovatively parameterized scene objects as feature vectors (including position, size, orientation, semantic category, and geometry), employing forward diffusion and reverse denoising processes to learn global scene layouts. This framework supports scene completion, object rearrangement, and text-conditioned generation. On this basis, to enhance the visual quality and representational capability of generated results, some works have begun to introduce visual modalities as guidance. CC3D \cite{bahmani2023cc3d} efficiently generated 3D scenes via feature field squeezing technology by leveraging bounding box-based semantic layouts; SpatialGen \cite{fang2025spatialgen} introduced a multi-view diffusion model to reconstruct 3D layouts into semantic 3DGS; and SceneCraft \cite{yang2024scenecraft} used multi-view images containing semantic categories and depth information as conditions to train a 2D diffusion model, and ultimately distilled a NeRF as the scene representation.

In complex multi-object scenes, how to achieve disentangled generation of objects and backgrounds has become a key challenge. Po et al. \cite{po2024compositional} introduced a joint control mechanism combining 3D bounding boxes with text prompts, implementing regional denoising strategies based on Voxel NeRF representations to enhance spatial controllability. DisCoScene \cite{xu2023discoscene} innovatively modeled scenes as semantic-agnostic 3D bounding box collections containing affine transformation parameters, thereby decoupling scene objects from backgrounds. Leveraging a GAN framework, it generated individual object radiance fields guided by layout priors, ensured photorealism through global-local discriminators, and supported object-level editing. Epstein et al. \cite{epstein2024disentangled} adopted a similar idea: by independently optimizing object representations before learning spatial affine transformation parameters, they achieved semantically constrained scene compositions. DreamDissector \cite{yan2024dreamdissector} further introduced neural category fields to decompose NeRF density fields into category-specific sub-NeRFs. Enhanced by deep concept mining for disentanglement accuracy, it converted sub-NeRFs into DMTet-structured meshes, establishing a novel representation framework for component-wise complex scene generation. In contrast to the aforementioned approaches, Zhang et al. \cite{zhang2021fast} discarded complex generative models and instead learned spatial relationship priors combined with template optimization. This approach ensures layout rationality while achieving real-time generation efficiency within seconds.

Compared with indoor environments, the generation of large-scale outdoor scenes faces challenges including high complexity of 3D spatial structures, vast scene scale, and scarcity of real-world datasets. To address these challenges, recent research has focused on leveraging layout and geometric knowledge provided by bird's-eye view (BEV) maps for outdoor scene generation. SceneDreamer \cite{chen2023scenedreamer} pioneered BEV representation for scene structures. Based on a GAN framework, it generated photorealistic unbounded 3D natural landscapes from random simplex noise and style codes. Both CityDreamer \cite{xie2024citydreamer} and GaussianCity \cite{xie2025generative} focused on urban landscape generation rather than wilderness scenes, adopting BEV representations with divergent technical implementations. Concretely, the former employed a decoupled generation strategy, decomposing cities into three independent modules: unbounded layouts, background environments, and building instances, ultimately fused via a compositor. The latter further converted BEV into compact point cloud representations, significantly reducing memory consumption and achieving nearly 60× performance improvement over CityDreamer. Addressing infinite scene expansion demands, BerfScene \cite{zhang2024berfscene} introduced BEV maps as layout priors, leveraging an equivariant U-Net architecture with low-pass filters and dynamic padding strategies to achieve seamless local scene stitching, effectively overcoming spatial boundary constraints.

Notably, the introduction of LLMs and VLMs marks the shift of scene generation toward the agent planning paradigm. LayoutGPT \cite{feng2023layoutgpt} leveraged in-context learning to transform the layout generation task into a CSS style code generation problem, and stimulated the planning capability of LLMs with a small number of examples to directly output numerical scene parameters. Similar to LayoutGPT, studies including \cite{zhang2024towards,gao2024graphdreamer, zhou2024gala3d, li2024dreamscene} extensively exploit LLMs' semantic comprehension capabilities to extract critical scene elements from textual prompts. Specifically, SceneWiz3D \cite{zhang2024towards} employed LLMs to disentangle objects and environments, representing them via DMTet and NeRF respectively. It automatically configured scene layouts through particle swarm optimization and optimized scenes via SDS with perspective RGB and panoramic RGBD views. GALA3D \cite{zhou2024gala3d} generated scene layouts based on LLM-parsed object topology relationships. Building upon instance Gaussian distributions from MVDream \cite{shi2023mvdream}, it introduced adaptive geometry control modules to refine shape features and optimized entire scenes via scene-level diffusion priors. DreamScene \cite{li2024dreamscene} decomposed scene prompts into object and environment descriptions, initialized object representations with sparse point clouds from Point-E \cite{nichol2022point}, and refined 3DGS through multi-timestep sampling and CSD loss \cite{yu2023text}. Gaussian filtering and texture refinement were applied alongside distinct three-stage camera sampling strategies for indoor/outdoor scenarios to produce high-quality, globally consistent, and editable 3D scenes. SceneTeller \cite{ocal2024sceneteller} extracted object position and orientation information from natural language, constructed layouts, and then retrieved matching furniture models from a 3D model database. Subsequently, it adopted 3DGS to represent the generated scenes and integrated diffusion models to enable flexible scene style editing. Sun et al. \cite{sun2025hierarchically} leveraged LLMs to generate hierarchical text descriptions of scenes, cooperated with a hierarchy-aware Graph Neural Network to infer relative positions of objects, and solved layout problems through divide-and-conquer optimization. Furthermore, VLMs have been introduced to enhance the visual perception capability of generation. LayoutVLM \cite{sun2025layoutvlm} proposed to use rendered images with Visual Marks to assist the model in perceiving spatial depth, and combined a self-consistent decoding strategy to generate layouts with both semantic and geometric rationality. ImmerseGen \cite{yuan2025immersegen} was tailored for VR environments; it utilized VLM agents to analyze terrain features based on Semantic Grids and accurately placed lightweight geometric proxies, which effectively improved the realism and immersion of natural scene generation. In contrast, ArtiScene \cite{gu2025artiscene} innovatively introduced high-quality 2D images as layout intermediaries, and inferred 3D spatial layouts through depth estimation and mask extraction techniques, thereby avoiding the ambiguity of layout generation relying solely on text. Scenethesis \cite{ling2025scenethesis} combined the commonsense reasoning of LLMs and the spatial perception of large vision models, and iteratively refined object layouts and interaction relationships through SDF constraints. To address the challenge that single-step inference struggles to handle complex long instructions, recent studies have shifted toward agent frameworks featuring multi-agent collaboration and iterative optimization. Holodeck \cite{yang2024holodeck} used GPT-4 to convert complex scene descriptions into spatial constraints and employed a depth-first search strategy to find object placement positions for generating reasonable layouts. I-Design \cite{ccelen2024design} simulated human design teams and constructed a multi-agent system including designers and engineers, which collaboratively generated scene graphs and solved layout problems through multi-round dialogue and backtracking mechanisms. Similarly, PhiP-G \cite{li2025phip} designed multiple agents: the keyword extraction agent constructed scene graphs from text descriptions, the generation agent was responsible for rapidly producing 3D assets, the classification agent accurately matched physical relationships between objects to ensure contact and fitting, and the supervision agent optimized the entire scene by evaluating layouts from multiple perspectives. SceneWeaver \cite{yang2025sceneweaver} introduced an agent framework with "reasoning-action-reflection" capabilities, which iteratively refined layouts by continuously invoking tools. For robot manipulation tasks, MesaTask \cite{hao2025mesatask} designed a Chain of Thought for spatial reasoning to infer the manipulation dependencies between objects. For large-scale urban scenes, Yo’City \cite{lu2025yo} simulated the hierarchical logic of urban planning: it generated regional grids via a global planner and then filled in architectural details with local designers, achieving cross-scale generation from macro to micro levels.

Furthermore, as an intermediate representation with greater semantic depth, the explicit node and edge structure of scene graphs can effectively handle complex semantic and spatial relationships between objects. Representative works such as Graph-to-3D \cite{dhamo2021graph} directly generated object layout bounding boxes and corresponding 3D shapes from scene graphs. CommonScenes \cite{zhai2023commonscenes} employed a dual-branch generation framework where a VAE predicts scene layouts while a latent diffusion model jointly generates diverse 3D object shapes, ensuring global consistency and local relational rationality. Subsequent studies, EchoScene \cite{zhai2024echoscene} and MMGDreamer \cite{yang2025mmgdreamer} proposed dual-branch diffusion models for shape and layout generation. The former further supports dynamic editing of nodes and edges in scene graphs, while the latter enhances scene graph expressiveness through multimodal node and relation prediction, enabling finer control over scene layouts and object shapes. To address attribute confusion and guidance collapse in complex scenes for existing text-to-3D methods, GraphDreamer \cite{gao2024graphdreamer} converted text into scene graphs, combining node-level disentangled modeling with edge relationship constraints. This approach optimizes SDF representations through SDS loss and geometric constraints to generate composable 3D scenes. Similarly, leveraging scene graphs to enhance the physical plausibility of generated content, LayoutDreamer \cite{zhou2025layoutdreamer} introduced an energy-based physical optimization module. By calculating physical potential energies such as gravity, support, and penetration, it finely adjusted the scene graph-based Gaussian sphere layout, ensuring that the generated scenes are not only semantically accurate but also conform to physical laws. The structured characteristics of scene graphs also enable their outstanding performance in instruction-based editing tasks. InstructScene \cite{lin2024instructscene} proposed a semantic scene graph-guided editing framework, which utilized GCN to encode semantic changes in scene graphs and guided an autoregressive Transformer to update layouts. This achieves precise control over object removal, insertion, or attribute modification within scenes. For visual guidance and cross-modal generation, Imaginarium \cite{zhu2025imaginarium} adopted an "image-as-layout" idea. It first generated high-quality 2D images as visual blueprints, extracted scene graph structures from these blueprints, and then used optimization algorithms to precisely align retrieved 3D assets with the perspective and semantic layout of the images. ScenePainter \cite{xia2025scenepainter} employed Scene Concept Graphs to align semantic relationships between objects, thereby guiding texture repainting under single-image conditions and preventing semantic drift during multi-view generation. Moreover, the application scope of scene graphs has been extended to large-scale outdoor scenes. Liu et al. \cite{liu2025controllable} designed an interactive system to convert sparse scene graphs into dense BEV embedding features, guiding a pyramid discrete diffusion model to generate voxelized urban and driving scenes. This breaks through the limitation that scene graphs were previously only applicable to small-scale indoor scene generation.

\subsection{2D Prior-based Generation}

The 2D prior-based scene generation paradigm has emerged as a dominant research direction. The core idea of such methods is to lift the rich knowledge embedded in images or large-scale pre-trained diffusion models into 3D space via spatial constraints.

Early explorations mainly focused on constructing scenes through an iterative "generation-fusion" process. Typically starting from sparse views, these methods leverage image-to-depth models to obtain geometric information, generate novel views by sampling camera poses, complete missing regions with inpainting, and finally fuse the results into a coherent 3D representation. As an early work in this domain, Shi et al. \cite{shi20223d} proposed a dual-path generator architecture based on GANs to simultaneously synthesize RGB images and depth maps for indoor scenes, and designed a switchable discriminator to enhance 3D consistency. GAUDI \cite{bautista2022gaudi} learned the latent distribution of 3D scenes via a DDPM \cite{ho2020denoising}, enabling unconditional and conditional 3D scene generation, yet it is constrained by low efficiency in both inference and rendering. NeuralField-LDM \cite{kim2023neuralfield} adopted a hierarchical latent diffusion framework to encode RGB images and camera poses, and learned the 3D scene distribution in the form of voxel grids. RGBD2 \cite{lei2023rgbd2} leveraged intermediate meshes for rendering and fusion, and employed a diffusion model for incremental view refinement to construct 3D indoor scenes. Text2Room \cite{hollein2023text2room} and SceneScape \cite{fridman2023scenescape} both adopted similar strategies, which iteratively generate and fuse meshes by utilizing monocular depth estimation and view refinement. To address the issue of geometric consistency, Text2NeRF \cite{zhang2024text2nerf} established a joint framework combining text-to-image diffusion models with NeRF, synthesizing multi-view consistent 3D scenes through progressive inpainting strategies. Invisible Stitch \cite{engstler2025invisible} focused on resolving seam artifacts in iterative generation, maintaining geometric coherence through a depth map refinement model and a self-training strategy. In the field of long-sequence scene generation, WonderJourney \cite{yu2024wonderjourney} introduced LLMs to generate coherent scene description sequences, which drive the visual module to produce a series of connected 3D scenes, enabling cross-style perpetual roaming. 3DGS has attracted attention from several works due to its efficient rendering capability. LucidDreamer \cite{chung2023luciddreamer} transformed multimodal inputs into multi-view aligned image collections, constructed initial point clouds via depth estimation, and optimized 3DGS representations. RealmDreamer \cite{shriram2024realmdreamer} leveraged a 2D inpainting model as a denoising prior to distill 3D Gaussians, and integrated a depth diffusion model to refine geometric structures. The latest work WonderWorld \cite{yu2024wonderworld} tackled the inefficiency of iterative generation by adopting layered Gaussian surfaces for initialization and fast optimization. It further took scene visible depth as a constraint to mitigate geometric seam artifacts, achieving second-level generation of interactive 3D scenes. Additionally, for texture generation given predefined scene geometry, SceneTex \cite{chen2024scenetex} proposed a multi-resolution texture field and a cross-attention decoder. It iteratively optimized the texture field during multi-view rendering with the aid of VSD \cite{wang2023prolificdreamer}, yielding high-quality textures with consistent stylistic features across all views.

\begin{figure}
    \centering
    \vspace{-0.4cm}
    \includegraphics[width=\textwidth]{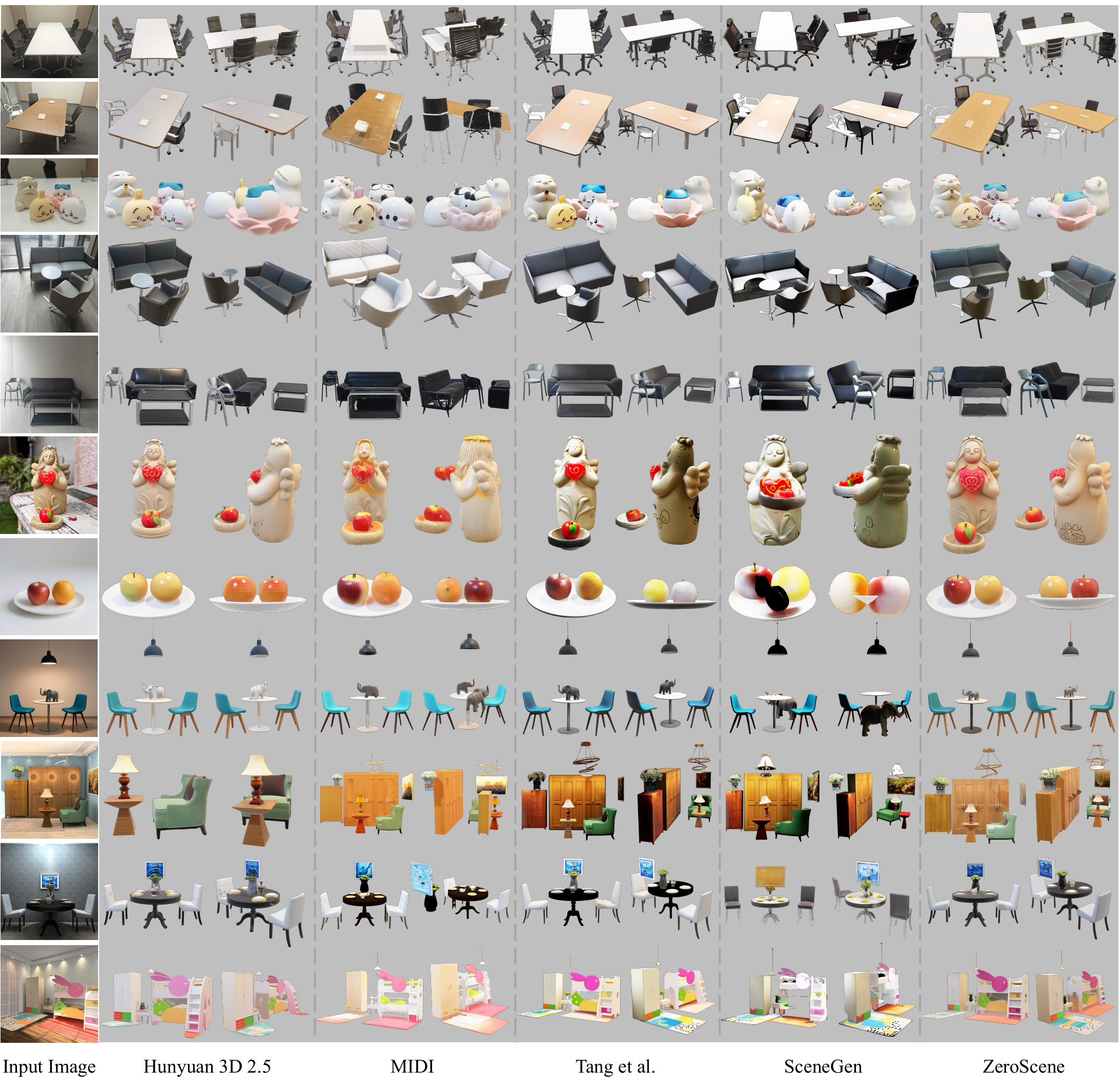}
    \vspace{-0.8cm}
    \caption{Qualitative comparison of single image-to-3D scene generation approaches.}
    \vspace{-0.6cm}
    \label{comparison_layout_methods}
\end{figure}

Compared to the stitching of multiple single-view images, panoramas often provide more comprehensive and geometrically consistent global priors, and thus are widely adopted as intermediate proxies for scene generation. RoomDreamer \cite{song2023roomdreamer} generated 360° panoramic images based on text prompts and geometric guidance, while jointly optimizing the geometry and textures of 3D meshes. GenRC \cite{li2024genrc} directly synthesized cross-view consistent panoramic RGBD images using a pre-trained diffusion model and a custom E-Diffusion technique, thereby completing scene geometry and texture. DreamCube \cite{huang2025dreamcube} proposed a multi-plane synchronization mechanism to generate seamless RGBD cubemaps for reconstructing high-quality scenes. For the lifting of panoramic-to-3D conversion, ControlRoom3D \cite{schult2024controlroom3d} employed latent diffusion models to generate panoramas with depth estimation for layout alignment, iteratively refining meshes. DreamScene360 \cite{zhou2024dreamscene360} and SceneDreamer360 \cite{li2024scenedreamer360} both adopted 3DGS as the representation. The former directly lifted 2D panoramic images to 3DGS, while the latter combined enhanced panoramic image generation with multi-view projection, initializing point clouds via monocular depth estimation and training 3DGS. Also based on 3DGS, FastScene \cite{ma2024fastscene} generated panoramic images and depth maps through diffusion models, realizing fast scene reconstruction with multi-view projection technology; HoloDreamer \cite{zhou2024holodreamer} proposed a two-stage panoramic reconstruction strategy to improve the completeness of 3D Gaussians. To address the occlusion problem of single panoramic images, LayerPano3D \cite{yang2025layerpano3d} decomposed panoramic images into multiple layers using depth information and performed inpainting to complete missing regions, enabling large-scale roaming. Similarly, HunyuanWorld 1.0 \cite{team2025hunyuanworld} utilized an agent-based pipeline to decompose the world into semantic layers such as sky, background, and objects, reconstructing and combining them hierarchically. Schwarz et al. \cite{schwarz2025recipe} adopted a similar approach, generating panoramic images via in-context learning and completing occluded areas. PERF \cite{wang2024perf} trained NeRF from single panoramic images, resolving multi-view geometric conflicts using collaborative RGBD inpainting and a progressive inpainting-and-erasing strategy. OmniX \cite{huang2025omnix} constructed a general framework based on Flow Matching models, unifying the perception and prediction of panoramic image generation and PBR material properties through cross-modal adapters, thereby building 3D scenes that support physically based rendering.

Unlike the aforementioned methods that aim to reconstruct the overall 3D scene from image collection or panoramic view, to address the challenges of severe inter-object occlusion and cluttered layout in complex scenes, an emerging trend is to decompose the scene into independent objects and layouts for separate modeling, i.e., achieving 3D reconstruction through parsing and decoupled reconstruction of a single image. Fig. \ref{comparison_layout_methods} and Table \ref{comparison_scene_generation} compare the qualitative and quantitative performance of various single-view 3D scene generation methods. Lay-A-Scene \cite{rahamim2024lay} generated scene graphs via personalized fine-tuning of Stable Diffusion \cite{rombach2022high} and inferred the 3D layout of objects in reverse. REPARO \cite{han2025reparo}, CAST \cite{yao2025cast}, Tang et al. \cite{tang2025towards}, ZeroScene \cite{tang2025zeroscene}, and TabletopGen \cite{wang2025tabletopgen} adopted a "decomposition-reconstruction-optimization" strategy: they first decomposed a single image into independent objects for individual reconstruction, and then optimized the spatial layout of objects using techniques such as differentiable rendering \cite{han2025reparo}, alignment modules with physical awareness \cite{yao2025cast}, projection loss \cite{tang2025towards, tang2025zeroscene}, and pose-scale alignment algorithms \cite{wang2025tabletopgen}. Following the same paradigm, ComboVerse \cite{chen2024comboverse} and Gen3DSR \cite{ardelean2025gen3dsr} leveraged space-aware SDS loss or cross-modal completion techniques to refine object layout and integrity. HiScene \cite{dong2025hiscene} first generated an isometric view as an overview, parsed the objects, and then used a video diffusion model for completion and layered reconstruction. For zero-shot reconstruction, Zhou et al. \cite{zhou2024zero} proposed a depth prior assembly framework, which integrated existing segmentation, completion, and depth estimation models to tackle occlusion and layout issues in scene reconstruction; Diorama \cite{wu2025diorama} introduced a retrieval mechanism that decomposed images via perception models and optimized CAD model layouts using geometric-semantic constraints.

\begin{table}
    \centering
    \vspace{-0.3cm}
    \caption{Quantitative comparisons of single image-to-3D scene generation methods. We employ Chamfer Distance (CD) and F-Score (F-S) for geometric evaluation, and CLIP/DINOv2 similarity for visual quality assessment. All metrics are calculated at both object (-O) and scene (-S) levels. We report average values for these metrics, alongside an analysis of instance separability and background handling.}
    \vspace{-0.3cm}
    \setlength{\tabcolsep}{2pt} 
    \renewcommand{\arraystretch}{1.05} 
    \resizebox{\linewidth}{!}{
    \begin{tabular}{lcccccccccc}
    \toprule
    \multirow{2}{*}{\centering \makecell{\textbf{Method}}} &  
    \multicolumn{4}{c}{\textbf{Geometric Metrics}} & 
    \multicolumn{4}{c}{\textbf{Visual Metrics}} & 
    \multirow{2}{*}{\centering \makecell{\textbf{Separable} \\ \textbf{Assets}}} &
    \multirow{2}{*}{\centering \makecell{\textbf{Background} \\ \textbf{Handling}}} \\
    \cmidrule(lr){2-5} \cmidrule(lr){6-9}
     & \textbf{CD-O}$\downarrow$ & \textbf{CD-S}$\downarrow$ & \textbf{F-S-O}$\uparrow$ & \textbf{F-S-S}$\uparrow$ &
     \textbf{CLIP-O}$\uparrow$ & \textbf{CLIP-S}$\uparrow$ & \textbf{DINO-O}$\uparrow$ & \textbf{DINO-S}$\uparrow$ & \\
    \midrule
    Hunyuan3D 2.5 \cite{lai2025hunyuan3d} & - & 0.0226 & - & 72.43 & - & 0.876 & - & 0.837 & \usym{2717} & \usym{2717} \\
    MIDI \cite{huang2025midi} & 0.0409 & 0.0384 & 42.76 & 65.58 & 0.814 & 0.841 & 0.782 & 0.819 & \usym{2714} & \usym{2717} \\ 
    Tang et al. \cite{tang2025towards} & 0.0267 & 0.0282 & 54.24 & 70.64 & 0.805 & 0.839 & 0.852 & 0.835 & \usym{2714} & \usym{2717} \\ 
    SceneGen \cite{meng2025scenegen} & 0.0223 & 0.0416 & 63.63 & 67.17 & 0.835 & 0.827 & 0.846 & 0.812 & \usym{2714} & \usym{2717} \\ 
    ZeroScene \cite{tang2025zeroscene} & 0.0163 & 0.0137 & 79.35 & 83.21 & 0.908 & 0.913 & 0.886 & 0.893 & \usym{2714} & \usym{2714} \\ 
    \bottomrule
    \end{tabular}
    }
    \vspace{-0.6cm}
    \label{comparison_scene_generation}
\end{table}

To break through the speed bottleneck of optimization-based methods, recent research has shifted toward training end-to-end feed-forward models that directly predict scene parameters from inputs, enabling second-level generation. Prometheus \cite{yang2025prometheus} adopted a two-stage approach: it first trained a GS-VAE to compress multi-view data, then trained a multi-view diffusion model to generate RGB-D latent variables, which were decoded into pixel-aligned 3D Gaussians. SplatFlow \cite{go2025splatflow} presented an integrated framework consisting of a multi-view rectified flow model and a Gaussian decoder, supporting both generation and editing. Bolt3D \cite{szymanowicz2025bolt3d} employed a DiT-based latent diffusion model to directly regress multi-view 3D Gaussian parameters from input images without test-time optimization. Targeting the complexity of multi-object scene generation, Dahnert et al. \cite{dahnert2024coherent} formulated scene reconstruction as a conditional diffusion process, which denoised the 3D poses and shape latent codes of all objects simultaneously via intra-scene attention mechanisms to ensure global consistency. MIDI \cite{huang2025midi} also extended single-object generation models by introducing a multi-instance attention mechanism, enabling simultaneous generation of multiple objects with coordinated spatial layouts in a single feed-forward pass. SceneGen \cite{meng2025scenegen} and SAM 3D \cite{chen2025sam} extracted all objects from a single image, and predicted the geometry, texture, and relative positions of objects concurrently through feed-forward networks. In addition, for domain-specific generation tasks, Ran et al. \cite{ran2024towards} proposed a multi-modal conditional LiDAR diffusion model that directly generated point cloud data for autonomous driving scenes via curve-aware compression; Sat2City \cite{hua2025sat2city} designed a cascaded latent-space diffusion model that generated 3D cities in the form of sparse voxel grids directly from satellite heightmaps as conditions.

\subsection{Rule-driven Modeling}

Rule-driven modeling is a technique for automatically creating 3D models and textures based on predefined rules, parametric systems, and mathematical functions. This approach enables controllable generation of complex and diverse content (e.g., terrains, buildings, vegetation) through finite rule sets, with its advantage lying in rapid and efficient content production without reliance on large-scale data training. However, designing low complexity yet highly robust generation rules remains a primary challenge in this field. The academic community has conducted systematic research and achieved significant progress in this domain.

Early studies laid the theoretical foundation based on grammar and geometric algorithms. L-system \cite{lindenmayer1968mathematical} served as a formal language system for simulating biological growth processes. It generated intricate structures through recursive string replacement rules and has been widely applied in plant modeling and fractal geometry, establishing itself as a foundational tool for procedural modeling. Parish et al. \cite{parish2001procedural} expanded on L-system to develop the CityEngine system, which pioneered rule-driven generation of roads, parcels, and buildings for urban scenes using minimal statistical and geographical input data. Building upon this work, M\"{u}ller et al. \cite{muller2006procedural} introduced CGA shape, enabling efficient generation of high-detail 3D cities from simple volumetric models through extended shape grammars and context-sensitive rules. Their implementation integrated CGA shapes into the CityEngine framework using C++. Lipp et al. \cite{lipp2011interactive} and Vanegas et al. \cite{vanegas2012procedural} focused on urban layout generation. The former combined hierarchical systems with graph-cut algorithms for iterative layout merging, integrating procedural generation with interactive editing to ensure layout validity. The latter proposed a compositional approach by independently generating and assembling urban parcels. To lower the barrier of rule design, Talton et al. \cite{talton2011metropolis} introduced Markov Chain Monte Carlo techniques for probabilistic inference to automatically optimize rules. In contrast, recent work by Merrell \cite{merrell2023example} proposed a graph grammar-based inverse modeling method, which automatically infers rules by analyzing the local similarity of input samples to generate new models with similar styles. Moreover, VoxCity \cite{fujiwara2025voxcity} automated the generation of semantic 3D urban models for environmental simulation by integrating and voxelizing publicly available geospatial data (such as building heights, land cover, etc.).

With the rise of language and vision models, modern procedural modeling increasingly follows a pipeline of 3D feature extraction from existing tools to generate rules/grammars, which then drives asset creation. Specifically, 3D-GPT \cite{sun20233d} parsed textual inputs via LLMs to select procedural functions from the Infinigen library and generated executable Blender Python scripts \cite{blender2018}, enabling dynamic editing and photorealistic rendering of natural scenes. SceneX \cite{zhou2024scenex} integrated 172 procedural modules and 11,284 static 3D assets with standardized APIs, combining LLM-driven planners for scene decomposition and asset placement to create controlled natural scenes and unbounded cities. CityX \cite{zhang2024cityx} employed procedural modules and multi-agent collaboration frameworks to generate photorealistic 3D urban environments from multimodal inputs (e.g., text descriptions, OSM files, semantic maps). In addition, Feng et al. \cite{feng2025text} utilized LLMs to parse scene features from text, derive 2D layouts, heightmaps, and textures through 3D layout generation modules, and dynamically adjust predefined CGA templates via LLMs to map 2D layouts and model features into fully editable 3D urban environments in CityEngine \cite{CityEngine2024}. Gumin et al. \cite{gumin2025procedural} further modeled scene generation as the process of writing Domain-Specific Language code, utilizing LLMs to generate program instructions and introducing an LLM-free search-based error correction mechanism to ensure the physical plausibility of layouts. In the field of interior design, SceneCraft \cite{hu2024scenecraft} innovatively adopted the paradigm of writing Blender Python code \cite{blender2018} for scene generation, and continuously revised the generated code through a feedback loop driven by VLMs, while also possessing the capability to learn new skill libraries from historical generation outcomes. RoomCraft \cite{zhou2025roomcraft}, on the other hand, integrated LLMs with the heuristic-based depth-first search algorithm and addressed complex furniture layout constraints via a conflict-aware positioning strategy, thereby achieving high-precision indoor scene generation.

To address the demands for large-scale scene generation, Raistrick et al. \cite{raistrick2023infinite} developed a mathematical rule-driven generation framework that parameterizes geometric shapes, textures, and materials to enable infinite combinations of natural assets (e.g., terrains, flora, fauna), seamlessly integrated with Blender \cite{blender2018}. Their subsequent work Infinigen Indoors \cite{raistrick2024infinigen} specialized in indoor scene generation, supporting dynamic mesh detail adjustment based on depth of field and user-defined complex layout constraints. To generate physically interactive scenes that provide training scenarios for embodied intelligence, ProcTHOR \cite{deitke2022} built a procedural generation framework based on multi-stage sampling. This framework is capable of producing a vast number of diverse and physically compliant indoor residential environments, which has significantly advanced the training of intelligent agents.

To break through the geometric limitations of rule-only generation, recent studies have begun to explore hybrid paradigms combining procedural modeling with neural rendering or generative models. Proc-GS \cite{li2024proc} incorporated procedural rules into the training process of 3DGS, leveraging procedural code to manage the distribution and variance of base assets, thus enabling architectural generation with both high-quality rendering performance and flexible editing capabilities. BuildingBlock \cite{huang2025buildingblock} proposed a two-stage hybrid approach: it first employed a diffusion model to generate the volumetric block layout of buildings, then used LLMs to enrich semantic rules and drive procedural modeling. This approach not only maintains structural rationality but also enhances the geometric details and diversity of the generated results.
\section{Challenge and Future work} \label{section6}

Benefiting from the emergence of novel 3D representations and advances in generative models, 3D generation methods have achieved significant technical breakthroughs. However, substantial obstacles persist when it comes to effectively applying the generated content to downstream tasks. This section discusses the remaining challenges in this realm and identifies potential future research directions.

\begin{table}
  \centering
  \vspace{-0.3cm}
  \caption{Classification of common datasets according to the data types. We have documented the size of each dataset and the number of object categories it contains, which are connected by the symbol "-".}
  \vspace{-0.3cm}
  \resizebox{\textwidth}{!}{
  \begin{tabular}{lcccc}
    \toprule
    \textbf{Datasets} & \textbf{Data type} & \textbf{Size} & \textbf{Categories} & \textbf{Year}\\
    \midrule
    ModelNet \cite {wu20153d} & 3D Model & 127K & 662 - Objects & 2014 \\
    ShapeNetCore \cite{chang2015shapenet} & 3D Model & 51K & 55 - Objects & 2015 \\
    PartNet \cite{mo2019partnet} & 3D Model & 26.6K & 24 - Objects & 2018 \\
    GSO \cite{downs2022google} & 3D Model & 1K & 17-Household Items & 2022\\
    Objaverse \cite{deitke2023objaverse} & 3D Model & 800K & Objects & 2022 \\
    Objaverse-XL \cite{deitke2023objaversexl} & 3D Model & 10.2M & Objects & 2023\\
    3D-Front \cite{fu20213dfront} & 3D Model + Indoor Layout & 13.1K(Models) + 18.9K(Rooms) & 31 - Indoor Scenes & 2020\\
    InternScenes \cite{zhong2025internscenes} & 3D Model + Indoor Layout & 1.96M(Models) + 40K(Scenes) & 288-Objects + 15-Indoor Scenes & 2025\\
    IL3D \cite{zhou2025il3d} & 3D Model + Indoor Layout & 29K(Models) + 27K(Layouts) & 18 - Houses & 2025\\
    3D-Future \cite{fu20213dfuture} & 3D Model + RGB Image & 16.5K(Models) + 20.2K(Images) & 34 - Furniture + Indoor scenes & 2020 \\
    MVImgNet \cite{yu2023mvimgnet} & RGB Image & 6.5M & 238 - Objects &  2023\\
    ScanNet \cite{dai2017scannet} & RGBD Image & 2.5M & Indoor Scenes & 2017\\
    HyperSim \cite{roberts2021hypersim} & RGBD Image & 77.4K & 461 - Indoor Scene & 2020 \\
    HouseLayout3D \cite{bieri2025houselayout3d} & RGBD Image + Indoor Layout & 26K(Images) + 317(Houses) & House structure & 2025 \\
    CO3D \cite{reizenstein2021common} & Video & 19K & 50 - Objects & 2021 \\
    uCO3D \cite{liu2025uncommon} & Video & 170K & 1000 - Objects & 2025 \\
    Text2Shape \cite{chen2019text2shape} & 3D Model - Text pairs & 15K(Models) + 75K(Text) & 2 - Objects &  2018 \\
    Text2Shape++ \cite{fu2022shapecrafter} & 3D Model - Text pairs & 369K & Objects & 2022 \\
    Point-E \cite{nichol2022point} & 3D Model - Text pairs & $>$1M(Models) + 120K(Text) & Objects & 2022 \\
    \bottomrule
  \end{tabular}}
  \vspace{-0.6cm}
  \label{3d_dataset}
\end{table}

\subsection{Datasets}

Beyond procedural modeling, which relies on predefined rules and grammars, modern approaches based on deep generative models are inherently data-driven, and their performance is highly dependent on the scale, quality, and diversity of training data. However, constructing large-scale 3D asset datasets still faces severe challenges. Compared with easily accessible 2D images on the Internet, the acquisition of 3D data often requires expensive professional scanning equipment or time-consuming manual modeling, resulting in extremely high data acquisition costs. This bottleneck has led to early public datasets generally suffering from limited object category coverage and low geometric complexity, which restricts the generalization ability of generative models. To address these challenges, the research community has built various types of datasets, as shown in Table \ref{3d_dataset}, which can be divided into three categories according to data modalities: 3D object and scene model datasets, 3D model-text paired datasets, and 2D image/video datasets.

Although many studies \cite{poole2022dreamfusion, chung2023luciddreamer, yu2023text, wang2023prolificdreamer, ma2024scaledreamer, zhang2024text2nerf, liang2024luciddreamer, yu2024wonderworld} have promoted the development of 3D generation methods based on 2D prior knowledge, the establishment of large-scale, high-quality 3D datasets remains of irreplaceable value. On one hand, such datasets can provide reliable benchmarks for training and validation of deep generative model-based 3D generation algorithms. On the other hand, standardized datasets will facilitate quantitative evaluation and comparative studies within the 3D generation research community.

\subsection{Evaluation Metrics}

Establishing a quantitative evaluation system for assessing the quality of 3D content generation remains one of the major challenges in the field of 3D generation. Current evaluation metrics are broadly classified into two categories: objective and subjective measurements.

Objective evaluation primarily focuses on quantifiable metrics. Traditional image quality assessment methods such as PSNR, SSIM, and LPIPS, along with generative adversarial network metrics including FID and KID, are employed to evaluate the rendered outputs of generated content. However, these 2D projection-based evaluation methods struggle to comprehensively capture the geometric consistency and diversity characteristics of 3D content. For geometric similarity measurement, metrics such as Chamfer Distance (CD) and Intersection over Union (IoU) assess geometric accuracy by quantifying point set or voxel differences between generated shapes and reference ground truth. To more comprehensively evaluate geometric quality, the F-score has been widely adopted. By integrating the precision and recall of point cloud matching, it addresses the limitation of the CD being sensitive to outliers. Furthermore, for the assessment of complex 3D scene generation, mere evaluation of rendering quality and geometric overlap detection is insufficient to fully measure physical plausibility. Existing physical evaluation methods are mostly confined to bounding box-based static collision detection, which overlooks the passability and functionality of scene layouts. In view of this, introducing navigation graph-based reachability analysis and path planning simulation will serve as a viable direction to assess whether generated scenes possess reasonable movement lines and spatial logic. This is expected to further compensate for the deficiency of geometric metrics in functional evaluation.

In terms of semantic consistency and visual perception evaluation, methods based on pre-trained large models have become mainstream. The CLIP Score \cite{radford2021learning} is employed to quantify the semantic alignment between text and 3D content (or its rendered images). Complementing this, DINO similarity \cite{oquab2024dinov2} is increasingly adopted to assess the structural consistency and visual fidelity of generated content; it can capture the overall layout of objects and the correspondence of local features more robustly than pixel-level metrics. Traditional CLIP Score and DINO similarity are often insufficiently sensitive when handling complex spatial relationships and fine-grained attributes, and they lack interpretability. To address this issue, several approaches \cite{ge2024creating, yang2025sceneweaver} have begun to leverage VLMs as judges to conduct multi-dimensional evaluations of rendered images. Going a step further, LEGO-Eval \cite{hwangbo2025lego} proposed a tool-augmented automated evaluation framework. Instead of being confined to black-box visual scoring, this method enables large models to call API tools to extract exact spatial coordinates and attribute information from scene graphs, thereby performing multi-hop reasoning verification on complex natural language instructions (e.g., relative positions between objects, material details). This paradigm shift from perceptual similarity to logical verification significantly improves the accuracy and robustness of semantic alignment evaluation for complex 3D scenes. In addition, as a text-to-3D generation benchmark, T$^3$Bench \cite{he2023t3bench} comprises three categories of text prompts: single objects, single objects with surrounding environments, and multiple objects. By introducing an automated evaluation framework that addresses two critical dimensions: 3D subjective quality assessment and text alignment evaluation, this benchmark effectively enhances the comparability of generation methods. Building upon this framework, we present qualitative comparisons of various diffusion-based generation approaches, as shown in Table \ref{comparison_on_T3_bench}. 

Subjective evaluation predominantly relies on user studies, which integrate human geometric cognition and visual perception capabilities to holistically assess the quality and diversity of generated content. However, this approach is not only time-consuming and labor-intensive but also prone to evaluators' subjective preferences, thereby compromising the objectivity and reproducibility of assessment outcomes.

Currently, a unified evaluation standard system has yet to be established. The development of a multi-dimensional evaluation framework that simultaneously addresses geometric precision, physical and functional plausibility, and fine-grained semantic consistency holds urgent theoretical significance and practical value for advancing 3D content generation technologies.


\begin{wraptable}{r}{7.1cm}
  \centering
  \vspace{-0.5cm}
  \caption{Quantitative comparisons on $\text{T}^3$Bench \cite{he2023t3bench}. We report the Running Time and the average scores (mean of quality and alignment) across three prompt categories: S-O (single object), W/S (single object with surrounding environments), and M-O (multiple objects). "Avg." denotes the overall average score.}
  \vspace{-0.3cm}
  \resizebox{0.5\textwidth}{!}{
  \begin{tabular}{lccccc}
    \toprule
    \textbf{Methods} & \textbf{Time} $\downarrow$ & \textbf{S-O} $\uparrow$ & \textbf{W/S} $\uparrow$ & \textbf{M-O} $\uparrow$ & \textbf{Avg.} $\uparrow$ \\
    \midrule
    DreamFusion \cite{poole2022dreamfusion} & 30 mins & 24.4 & 24.6 & 16.1 & 21.7    \\
    Fantasia3D \cite{chen2023fantasia3d} & 45 mins & 26.4 & 27.0 & 18.5 & 24.0    \\
    Magic3D \cite{lin2023magic3d} & 40 mins & 37.0 & 35.4 & 25.7 & 32.7    \\
    ProlificDreamer \cite{wang2023prolificdreamer} & 240 mins & 49.4 & 44.8 & 35.8 & 43.3    \\
    MVDream \cite{shi2023mvdream} & 30 mins & 47.8 & 42.4 & 33.8 & 41.3    \\
    DreamGaussian \cite{tang2023dreamgaussian} & 7 mins & 19.8 & 14.1 & 10.9 & 14.9    \\
    GeoDream \cite{ma2023geodream} & 400 mins & 41.1 & 34.9 & 25.4 & 33.8    \\
    GaussianDreamer \cite{yi2024gaussiandreamer} & - & 54.0 & 48.6 & 34.5 & 45.7    \\
    \bottomrule
  \end{tabular}}
  \vspace{-0.4cm}
  \label{comparison_on_T3_bench}
\end{wraptable}

\subsection{Potential Research Direction}

3D generation, as a key technology in computer graphics and vision, holds significant application value in cutting-edge scenarios. Although contemporary 3D content generation methods have achieved remarkable progress in output quality, diversity, and efficiency, they still exhibit critical limitations regarding precise controllability, scalability for large-scale scenarios, temporal evolution of dynamic scenarios, and interaction-oriented physical plausibility. Consequently, the practical implementation of these generative outcomes in real-world applications poses formidable challenges.

Regarding controllability, current automated generation tools, particularly deep generative models, whether employing unconditional generation or conditional generation based on text/image inputs, demonstrate that existing architectures still struggle to fully reproduce the geometric details and appearance characteristics anticipated by users. In contrast, rule-driven procedural modeling offers excellent controllability, yet the complexity of designing rule systems and the difficulty of algorithmic implementation significantly raise the technical entry barrier. Further utilization of mature tools such as language and vision models to extract scene features for rule generation, thereby enabling inverse procedural modeling \cite{guo2020inverse, zhai2024cropcraft, dai2025meshcoder}, represents a potential future research direction. 

Furthermore, the generation and extension of unbounded scenes warrant in-depth discussion. The core objective lies in developing algorithms or models capable of generating infinitely extensible 3D scenes or seamlessly expanding and integrating existing finite scenes. Key challenges in this field primarily involve two aspects: First, explicit 3D representations struggle to address the storage and transmission demands of infinite scenes, necessitating research into implicit representation frameworks and efficient data compression methodologies. Second, it is imperative to ensure structural and textural consistency between extended regions and original scenes to prevent visual discontinuities. Addressing these challenges, some works \cite{chen2023scenedreamer, xie2024citydreamer, xie2025generative, zhang2024berfscene} have explored unbounded natural landscapes and urban environments. Emerging studies indicate that divide-and-conquer strategies combined with multi-scale modeling can effectively decompose scene complexity. Concretely, both LT3SD \cite{meng2025lt3sd} and WorldGrow \cite{li2025worldgrow} adopted a coarse-to-fine block-wise generation strategy to support the expansion of infinite indoor scenes. The former decomposed 3D scenes into a hierarchical latent tree structure, where each layer contains a geometric volume and a latent feature volume for capturing low-frequency information and encoding high-frequency details, respectively. The latter was based on 3D Structured Latent Variables and leveraged a flow matching model to perform block-wise 3D scene completion and expansion. SynCity \cite{engstler2025syncity} employed a tile-based generation strategy, which elevated the generation capability of 2D diffusion models to the 3D space and enabled the stitching and synthesis of infinite cities. Liu et al. \cite{liu2024pyramid} investigated diffusion models for large-scale outdoor 3D scene synthesis. Inspired by pyramid-based multi-scale modeling, this approach decomposed scenes into multiple scales, each governed by an independent diffusion model. The system progressively refined local features based on coarse-grained scene layouts, utilizing scene partitioning to overcome GPU memory constraints and theoretically supporting infinite-scale scene generation. BlockFusion \cite{wu2024blockfusion} generated high-quality 3D scene geometry through tri-plane compression and latent space diffusion, achieving infinite scene expansion via extrapolation mechanisms. 

It is worth noting that the real world is not merely static, and dynamic 3D generation (i.e., 4D generation) serves as a crucial bridge connecting static geometry and physical interactions. To capture the temporal evolution of scenes, researchers have introduced dynamic NeRF and 4D Gaussian Splatting techniques. D-NeRF \cite{pumarola2021d} decomposed dynamic scenes into a static canonical space and a time-varying displacement field, enabling the learning of continuous motions of non-rigid objects from monocular videos. Duan et al. \cite{duan20244d} employed anisotropic 4D Gaussian spheres to represent dynamic scenes and projected the 4D manifold into 3D Gaussian distributions for each frame via a temporal slicing mechanism, thus achieving efficient dynamic novel view synthesis. In terms of generation paradigms, mainstream methods typically incorporate temporal priors through video diffusion models, transforming the generation task into a video-to-4D lifting process. For instance, 4Real-Video-V2 \cite{wang20254real} adopted a DiT architecture integrated with view-time attention, and directly regressed synchronized multi-view dynamic 3DGS through feed-forward networks. V2M4 \cite{chen2025v2m4} leveraged a native 3D mesh generation model \cite{xiang2025structured} for frame-by-frame reconstruction. By virtue of rigorous geometric-topological registration and texture optimization, it achieves the construction of topologically consistent 4D mesh assets from monocular videos.

After the generation of static and dynamic scenes has been addressed, endowing scenes with physical properties and interactive capabilities constitutes a critical step toward Embodied AI. Merely generating visual appearances can no longer meet the demands of downstream tasks, and research focus is shifting toward physical consistency and full-stack asset generation. PhysGen3D \cite{chen2025physgen3d} attempted to infer physical parameters from single images and integrated the Material Point Method particle approach, ensuring that the motions of generated objects comply with physical laws. RainyGS \cite{dai2025rainygs} integrated the accuracy of physical rain simulation with the efficient rendering capabilities of 3DGS. Through splash simulation and particle collision detection, it achieves efficient synthesis of physically credible rainy scenes in open environments. More comprehensively, EmbodiedGen \cite{wang2025embodiedgen} has built a large-scale generation toolkit tailored for embodied AI tasks, encompassing multiple modules such as text-to-3D, image-to-3D, articulated object generation, texture generation, scene generation and layout generation. This toolkit provides rich and fully functional 3D assets for the interaction between robots and simulation environments. These explorations demonstrate the potential of combining physical constraints with generative models. However, research on object/scene interaction and the modeling of physical properties for generated objects still requires deeper investigation.

Finally, in terms of scene visualization effectiveness, while neural representations \cite{mildenhall2021nerf} based on volume rendering achieve high-fidelity visual quality, their compatibility challenges with rasterization-based graphics pipelines hinder direct integration into traditional workflows like game engines. Emerging techniques like 3DGS have improved real-time performance, yet they still exhibit limitations in material expressiveness under complex illumination conditions. Furthermore, real-time rendering of unbounded scenes necessitates the integration of level-of-detail systems, view frustum culling, and asynchronous loading mechanisms to balance detail precision with computational resources. Future research could focus on the efficient generation of complex scenes while also exploring real-time rendering techniques.
\section{Conclusion}

This paper presents a comprehensive review on static 3D object and scene generation. We commence by investigating fundamental theoretical frameworks of 3D representations, analyzing the impacts of explicit, implicit, and hybrid representations on the quality of generative algorithms and computational efficiency. Subsequently, we systematically classify and summarize 3D object generation methods based on four categories of generative models. Expanding the scope to complex scene levels, we then elaborate in detail on the internal mechanisms and evolution paths of three categories of scene generation paradigms: layout-guided generation, lifting based on 2D priors, and rule-driven modeling. The survey concludes with an in-depth discussion of unresolved challenges in this field and proposes potential research directions for future exploration. Through this systematic examination, we aim to establish a structured technical reference framework for 3D content generation, providing both theoretical foundations and technical insights to facilitate subsequent research advancements.

\bibliographystyle{ACM-Reference-Format}
\bibliography{sample-base}



\end{document}